%% file: main.tex
\documentclass[sigconf, nonacm, pdfa]{acmart}

\newcommand\vldbdoi{XX.XX/XXX.XX}

\newcommand\vldbavailabilityurl{https://github.com/GraphRecommendation/gsampling}
\newcommand\vldbpagestyle{plain}

\input{preamble}

\input{glossaries}

\definecolor{ao(english)}{rgb}{0.0, 0.0, 0.0}
\newcommand{\changed}[1]{{\color{ao(english)}#1}}

\begin{document}

\title{
The Limits of Graph Samplers for Training Inductive Recommender Systems: Extended results
}

\author{Theis E. Jendal}
\affiliation{%
  \institution{Aalborg University}
}
\email{tjendal@cs.aau.dk}

\author{Matteo Lissandrini}
\affiliation{%
  \institution{University of Verona }
}
\email{matteo.lissandrini@univr.it}

\author{Peter Dolog}
\affiliation{%
  \institution{Aalborg University}
}
\email{dolog@cs.aau.dk}

\author{Katja Hose}
\affiliation{%
  \institution{TU Wien}
}
\email{katja.hose@tuwien.ac.at}

\begin{abstract}
\input{sections/00_abstract}
\end{abstract}

\maketitle

\pagestyle{\vldbpagestyle}

\ifdefempty{\vldbavailabilityurl}{}{
\vspace{.3cm}
\begingroup\small\noindent\raggedright\textbf{PVLDB Artifact Availability:}\\
The source code, data, and/or other artifacts have been made available at \url{\vldbavailabilityurl}.
\endgroup
}

\input{rules}

\RestoreAcronyms
\glsresetall
\section{Introduction}
\label{sec:introduction}
\input{sections/01_introduction}

\section{Background \& Preliminaries} 
\label{sec:problem_formulation}
\input{sections/02_problem_formulation}

\section{Related Work}
\label{sec:relatedWork}
\input{sections/03_related_work_v2}


\section{Methodology} 
\label{sec:metodology}
\input{sections/05_methodology}


\section{Experiments}
\label{sec:experiments}
\input{sections/06_experiments}

\section{Conclusion and future work}
\label{sec:conclusion}
\input{sections/99_conclusion}

\begin{acks}
  This research was partially funded by the Danish Council for Independent Research (DFF) under grant agreement no. DFF-8048-00051B and the Poul Due Jensen Fond (Grundfos Foundation). 
\end{acks}
\balance
\pagebreak

\bibliographystyle{ACM-Reference-Format}
\bibliography{references_short}

\clearpage
\appendix
\nobalance
\input{appendix/00_index}

\end{document}

%% file: preamble.tex
\usepackage{xcolor}

\usepackage{balance}

\usepackage{glossaries-extra}
\setabbreviationstyle[no-long]{short}  

\newcommand{\cautoref}[2]{\hyperref[#1]{\autoref*{#1}#2}}  

\usepackage[inline]{enumitem}  
\setlist{leftmargin=*}

\usepackage[english]{babel} 

\usepackage{array}
\usepackage{tabularray}
\usepackage{tabu}
\newcolumntype{R}{>{\setbox0=\hbox\bgroup}c<{\egroup}@{}}
\NewColumnType{R}{>{\setbox0=\hbox\bgroup}c<{\egroup}@{}}
\newcolumntype{A}[1]{>{\centering\arraybackslash}m{#1}} 
\newcolumntype{Z}{>{\setbox0=\hbox\bgroup}c<{\egroup}@{\hspace*{-\tabcolsep}}} 
\usepackage{pifont} 
\usepackage{multirow}
\usepackage{subcaption}
\usepackage{ragged2e}

\usepackage[export]{adjustbox}
\newlength{\imageheight}

\usepackage{algorithm}
\usepackage[noend]{algpseudocode}
\algnewcommand{\LineComment}[1]{\State \(\triangleright\) #1}
\algnewcommand{\Input}[1]{\State \textbf{input:} #1}
\algnewcommand{\Output}[1]{\State \textbf{output:} #1}

\definecolor{shade}{RGB}{119,122,138}



%% file: glossaries.tex
\newacronym{kg}{KG}{Knowledge Graph}
\newacronym{kge}{KGE}{Knowledge Graph Embedding}
\newacronym{ckg}{CKG}{Collaborative Knowledge Graph}
\newacronym{cg}{CG}{Collaborative Graph}

\newacronym{gnn}{GNN}{Graph Neural Network}
\newacronym{rs}{RS}{Recommender System}
\newacronym{cf}{CF}{Collaborate Filtering}
\newacronym{rnn}{RNN}{Recurrent Neural Network}
\newacronym{nlp}{NLP}{Natural Language processing}
\newacronym{bpr}{BPR}{Bayesian Personalized Ranking}
\newacronym{cnn}{CNN}{Convolutional Neural Network}
\newacronym{gcn}{GCN}{Graph Convolutional Network}
\newacronym{asha}{ASHA}{Asynchronous Successive Halving}
\newacronym{mlp}{MLP}{Multi-Layer Perceptron}

\newacronym{toppop}{TopPop}{Top Popular}
\newacronym{ppr}{PPR}{Personalized PageRank}
\newacronym{bprmf}{BPR-MF}{Bayesian Personalized Ranking for Matrix Factorization}
\newacronym{kgat}{KGAT}{Knowledge Graph Attention Network}
\newacronym{ae}{AE}{AutoEncoder}
\newacronym[category=no-long]{own}{GInRec}{Gated Inductive Recommendation}
\newacronym{mpnn}{MPNN}{Message Passing Neural Network}
\newacronym{ngcf}{NGCF}{Neural Graph Collaborative Filtering}
\newacronym{lgcn}{LGCN}{LightGCN}

\newacronym{ff}{FF}{Forest Fire}
\newacronym{rj}{RJ}{Random Jump}
\newacronym{rw}{RW}{Random Walk}
\newacronym{ts}{TS}{Temporal Sampling}
\newacronym{ps}{PS}{PinSAGE Sampling}

\newacronym{mr}{MR}{MindReader}
\newacronym{ml}{ML-20m}{MovieLens-20m}
\newacronym{mls}{ML-S}{MovieLens Subsampled}
\newacronym{ab}{AB}{Amazon-Book (2014)}
\newacronym{abs}{AB-S}{Amazon-Book Subsampled}
\newacronym{fm}{FM}{LFM-1b}
\newacronym{syn}{SYN}{Synthetic-1m}
\newacronym{yd}{YD}{Yelp Dataset}

\newacronym[longplural={Recommendable Entities}]{re}{RE}{Recommendable Entity}
\newacronym[longplural={Descriptive Entities}]{de}{DE}{Descriptive Entity}
\newacronym{sota}{SotA}{State-of-the-Art}

%% file: sections/00_abstract.tex
Inductive Recommender Systems are capable of recommending for new users and with new items thus avoiding the need to retrain after new data reaches the system. 
However, these methods are still trained on all the data available, requiring multiple days to train a single model, without counting hyperparameter tuning. 
In this work we focus on graph-based recommender systems, i.e., systems that model the data as a heterogeneous network.
In other applications, graph sampling allows to study a subgraph and generalize the findings to the original graph.
Thus, we investigate the applicability of sampling techniques for this task.
We test on three real world datasets, with three state-of-the-art inductive methods, and using six different sampling methods. 
We find that its possible to maintain performance using only $50\%$ of the training data with up to $86\%$ percent decrease in training time; however, using less training data leads to far worse performance. 
Further, we find that when it comes to data for recommendations, graph sampling should also account for the temporal dimension.
Therefore, we find that if higher data reduction is needed, new graph based sampling techniques should be studied and new inductive methods should be designed.
    

%% file: rules.tex
\newcommand{\verts}[1]{\mathcal{V}_{#1}}  
\newcommand{\edges}{\mathcal{E}}  
\newcommand{\users}{\mathcal{U}}  
\newcommand{\warmusers}{\mathcal{U}_w}  
\newcommand{\coldusers}{\mathcal{U}_c}  
\newcommand{\worrecs}{\mathcal{I}_{wor}}  
\newcommand{\wrrecs}{\mathcal{I}_{wr}}  
\newcommand{\feedback}{\mathcal{C}}  

\newcommand{\entities}{\mathcal{V}}  
\newcommand{\warmrecs}{\mathcal{I}_w}  
\newcommand{\coldrecs}{\mathcal{I}_c}  
\newcommand{\items}{\mathcal{I}}  
\newcommand{\recs}{\mathcal{I}}  
\newcommand{\descs}{\entities_{desc}}  
\newcommand{\labels}{\mathcal{L}}  
\newcommand{\relations}[1]{\mathcal{R}_{#1}}  
\newcommand{\real}{\mathbb{R}}  
\newcommand{\neighborhood}[1]{{\mathcal{N}_{#1}}}

\newcommand{\graph}[1]{\mathcal{G}_{#1}} 
\newcommand{\cg}{\graph{cg}}
\newcommand{\kg}{\graph{kg}}
\newcommand{\ckg}{\graph{ckg}}
\newcommand{\subgraph}[1]{#1'}

\newcommand{\user}{u}
\newcommand{\otheruser}{u'}

\newcommand{\cvert}{v}
\newcommand{\othervert}{v'}

\newcommand{\rec}{i}
\newcommand{\otherrec}{i'}

\newcommand{\entity}{{e}}
\newcommand{\otherentity}{{e'}}

\newcommand{\head}{h}
\newcommand{\relation}{r}
\newcommand{\tail}{t}

\newcommand{\ratio}{\alpha}
\newcommand{\probability}{p}
\newcommand{\ctime}{t}

\newcommand{\loss}[1]{\mathcal{L}_{#1}}

\newcommand{\interactions}[1]{\mathbf{I}_{#1}}
\newcommand{\cmatrix}[1]{\mathbf{\MakeUppercase #1}}
\newcommand{\iembedding}{\cmatrix{X}}

\newcommand{\ranking}{\mathbf{r}}
\newcommand{\cvector}[1]{\mathbf{\MakeLowercase #1}}
\newcommand{\hidden}[1]{\cvector{\entity}_{#1}}
\newcommand{\lemb}[3]{\cvector{#1}_{#2}^{(#3)}}
\newcommand{\lembtop}[3]{\cvector{#1}_{#2}^{(#3)\top}}

\newcommand{\ratingfunc}{R}
\newcommand{\Q}[1]{Q^{#1}}
\newcommand{\relmapping}{\phi}
\newcommand{\mse}{\text{MSE}}
\newcommand{\func}{\mathcal{F}}
\newcommand{\feature}{\mathcal{X}}
\newcommand{\sampler}{\mathcal{S}}

\newcommand{\preference}[1]{\leqslant_{#1}}
\newcommand{\predpref}[1]{\:\widehat{\preference{#1}}\:}
\newcommand{\topn}{top-$n$}
\newcommand{\pipe}{\bigm|}

%% file: sections/01_introduction.tex
\glspl{rs} are used in many applications, ranging from online retail stores to advertisement platforms. 
These systems utilize historic interactions between users and items to estimate future user behaviors, with the hypothesis that users with similar historical preferences will exhibit similar behavior in the future;  often referred to as \gls{cf}~\cite{jendal2023ginrec}.
Most approaches capture user preferences and item concepts as dense vector representations, called embeddings, within high-dimensional spaces, such that similar users and items have similar embeddings~\cite{jendal2023ginrec,wu2022inmo}.
To build these representations, deep neural networks learn vector representations assigned to all users and items often through dictionary encodings. 
These are called transductive techniques~\cite{jendal2023ginrec,wu2022inmo}.
This also means that, when a new user or item is added to the system, in theory they are required to re-train the model to compute the missing embeddings.

\begin{figure*}[!tb]
    \centering
    \includegraphics[width=.85\linewidth,trim=395 260 525 165, clip]{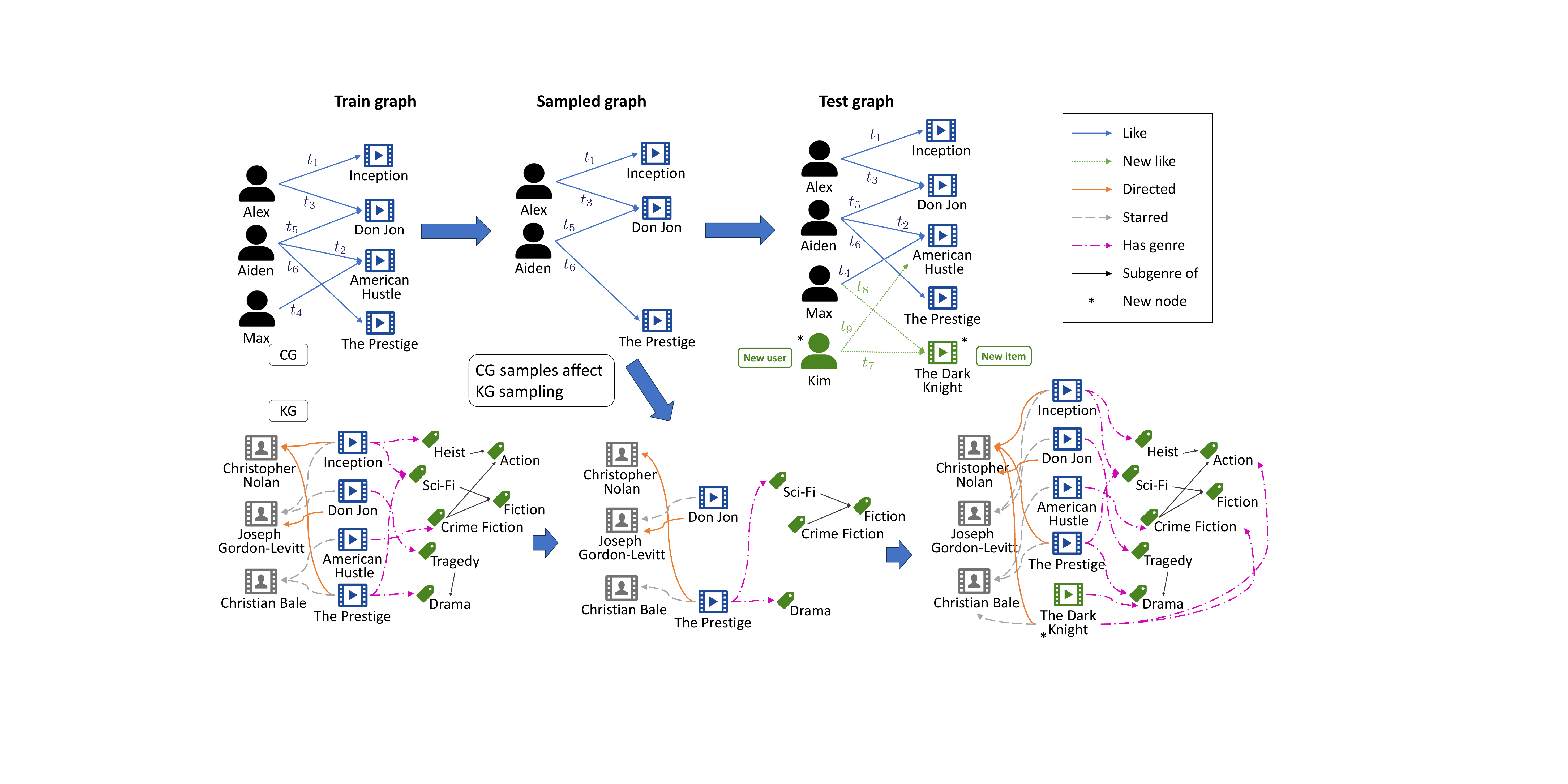}%
    \vspace{-5pt}
    \caption{Sampling example with illustration of sampling graph and the correlation between CG samples and KG samples.}
    \label{fig:sampling_example}
    \vspace{-5pt}
\end{figure*}

Recently, a lot of focus has been placed on inductive \glspl{rs} due to their ability to predict for unseen users and items~\cite{jendal2023ginrec,wu2022inmo,ying2018graphpinsage,zhang2019inductiveigmc,zhang2022geometricgimc}. 
These systems do not learn a unique vector for each user and item but instead learn to generate vectors based on their features and connections.
A transductive \gls{rs} would not be able to recommend ``The Dark Knight'' in \autoref{fig:sampling_example} as it is not present in the train graph.
In contrast, an inductive \gls{rs} can recommend items (and to users) absent during training but introduced at inference time~\cite{wu2022inmo,lee2019melu,sun2019bert4rec,wu2021towardsidcf,jendal2023ginrec}.
Among inductive methods, only a few can recommend effectively for both new users and items (see \autoref{table:methods}).
Yet, the training time of these inductive methods can be very slow, taking up to \textit{2 days to train} on a \gls{cg} with ${\sim}175$ thousand users and ${\sim}77$ thousand products.
Such long training times are particularly impactful for hyperparameter tuning, where multiple training cycles are often required.
Therefore, recent works study how to sample training data to decrease tuning time~\cite{garcia19samplinghpo,montanari22recommendationhpo}.
However, they focus on hyperparameter tuning, perform random sampling, and, most importantly, still require the methods to train on the full data afterwards.
This is particularly limiting if we consider that the graphs continuously evolve, with millions of items being added each day in some cases~\cite{liu2017pinterestrealworldrs}. 

In the past, graph sampling has been proven effective for studying important graph properties on a smaller scale~\cite{leskovec06samplinglargegraphs}.
Thus, in this paper, we are interested in studying whether it is possible to utilize graph sampling approaches to reduce the computational cost and, hence, the training time in the training step of graph-based \glspl{rs}.
Since inductive methods are capable of predicting for new users and items, we, in theory, do not require any retraining of the methods on the full dataset to be able to perform inference on it and thus to recommend new items or to new users. 

When performing graph sampling, current methods only sample within each batch, reducing batch forward propagation time but maintaining the number of batches in an epoch~\cite{liu2022sampling}. 
Therefore, they still go through all available graph data.
Instead, by subsampling the graph \textit{before} training, we effectively obtain a smaller graph structure and thus reduce the training data that needs to be processed.
We are interested in studying how to find a suitable subset of training nodes, that allows us to obtain an induced subgraph that is representative enough for the models to learn an inductive bias suitable for future recommendation.
The objective being to reduce training time with as little impact on final predictive quality as possible.
While the graph-sampling literature has proposed many different techniques~\cite{liu2022sampling,leskovec06samplinglargegraphs}, these techniques have not been studied with graph neural networks, the current de-facto standard architecture in \glspl{rs}.
Further, only one sampling method (node-based random sampling) and \gls{rs} (PinSAGE~\cite{ying2018graphpinsage}) has been tested directly on just a sub-sampled graph.
That is, the more established graph sampling techniques have yet to be tested in this domain.
We therefore study three state-of-the-art inductive \glspl{rs} on three real-world datasets using six graph sampling methodologies, including the sampling technique used in practice.
In summary, in this work we present:
\begin{enumerate*}
    \item The first extensive study of graph-based sampling prior to training for inductive recommender systems;
    \item A holistic evaluation of the limitations of current sampling methodologies and inductive \glspl{rs}; and
    \item A set of interesting research directions for the design of sampling techniques in inductive recommender systems.
\end{enumerate*}
\textbf{Our results demonstrate that:}
\begin{enumerate*}[label=\bfseries (\roman*)]
    \item It is possible to maintain good predictive performance by training on $50\%$ of the data while decreasing, in this way, the training time by up to $85\%$.
    \item Temporal sampling and user-based sampling perform best.
    \item For datasets with a high popularity bias, it is often enough to use $5\%$ of data for the system to perform well; and
    \item  with sampling ratios below $50\%$ existing sampling techniques and existing \glspl{rs} still struggle to maintain good performances; this raises the question of whether it is indeed possible to design more representative sampling algorithms and more robust learning approaches. 
\end{enumerate*}

%% file: sections/02_problem_formulation.tex
Similar to previous studies~\cite{jendal2023ginrec}, we consider \glspl{rs} using users, items, and positive interactions as input data. 
Further, we also allow for textual information and attributes attached to items. 
Formally, given a set of users $\users$ and a set of items $\recs$, we define an interaction matrix $\interactions{}{\in} \{0,1\}^{|\users|\times|\recs|}$, where $\interactions{\user\rec}{=}1$ if a user $\user{\in}\users$ has interacted with an item $\rec{\in}\recs$; otherwise $\interactions{\user\rec}{=}0$, i.e., the user has never interacted with the item. 
The interaction information can be structured as a bipartite graph, known as a \gls{cg}, where rating interactions appear as edges. 
Thus, the \gls{cg} can be defined as a $\cg{=}\langle \verts{cg},\relations{cg} \rangle$, where $\verts{cg} {=} \users \cup \recs$ are the users and items and $\relations{cg} {=} \left\{ (\user, \rec)| \interactions{\user\rec}{=}1 \right\}$. 
Furthermore, we define the mapping function $\func_\ctime{:}\users{\times}\recs\rightarrow\mathbb{N}^{\geq0}$, mapping all rating interactions (edges) to a natural number representing the time at which the rating was made, represented as $t_i$ in \autoref{fig:sampling_example}.
\changed{
The temporal aspect of ratings are important as trends and user interests change over time. 
The evaluation of \glspl{rs} should, therefore, take temporal information into account when constructing train, validation, and test sets. 
}

In addition, a \gls{kg}~\cite{wang2019kgat,jendal2023ginrec}, a heterogeneous graph containing entities and their semantic relations, is added to model descriptive information for items (see  \autoref{fig:sampling_example}). 
A \gls{kg} is a directed labeled multigraph defined as the triple $\kg{=}\langle \verts{kg}, \relations{kg}, \labels \rangle$ including nodes for both recommendable entities (items) and descriptive entities ($\descs{=}\verts{kg}\setminus\recs$).
Furthermore, the labels $\labels$ represent the semantic type of edges, s.t. the relationship can be defined as $\relations{kg}{\subseteq}\verts{kg}{\times}\labels{\times}\verts{kg}$.
In this model, the \gls{kg} does not represent the collaborative signal; thus we combine the \gls{kg} and \gls{cg} as a \gls{ckg}~\cite{wang2019kgat,jendal2023ginrec}, st, $\ckg{=}\langle \verts{ckg}, \relations{ckg}, \labels_{ckg}\rangle$, where $\verts{ckg}{=}\verts{kg}{\cup}\users$, $\relations{ckg}{=}\relations{kg}{\cup}\{(\user, likes, \rec)|\interactions{\user\rec}{=}1\}$, and $\labels_{ckg}{=}\labels \cup \{likes\}$.
We further include a feature function $\feature{:}\verts{kg}{\rightarrow}\real^{d}$ mapping each entity to a feature vector representing, for instance, the textual information for the node and the structure of $\kg$.

We treat the recommender objective as a ranking problem.
Thus, a recommender is a function $\ranking_{\user,\items}{=}\func_\theta(\feature, \interactions{}, \kg, \user)$ parametrized by learned parameters $\theta$ producing a ranking score for all items in $\items$ according to the inferred preferences of user $\user$. 
Thus, given the ranking 
$\ranking_\user$, it must hold that $\forall{\rec}, \otherrec{\in}\recs$, with $\rec{\neq}\otherrec$, we have that $\ranking_{\user,\rec}>\ranking_{\user,\otherrec}$ iff. the user $\user$ prefers item $\rec$ over $\otherrec$. 

Given the above data model and a sampling ratio $\alpha$, a sampling method $\sampler$ produces subgraphs $\subgraph{\cg}{\sqsubset}\cg$ and $\subgraph{\kg}{\sqsubset}\kg$ such that $|\subgraph{\cg}|{+}|\subgraph{\kg}| \leq \alpha{\cdot}(|\cg|{+}|\kg|)$. 
As in prior work~\cite{ying2018graphpinsage}, our goal is to train on the subgraphs $\subgraph{\cg}$ and $\subgraph{\kg}$ to learn the parameters for $\func_\theta$ and perform inference on the full graph.










%% file: sections/03_related_work_v2.tex



In the inductive setting, we have users and items not seen during training, for which the \gls{rs} should be able to make recommendations.
This capability is crucial for real-world applications where users and items are continuously added.
Further, it allows to train on a sub-graph while performing predictions for the entire graph.

\textbf{\emph{Inductive Recommender Systems.}}
There are multiple methods for inductive recommendation, using different techniques ranging from graph-based methods~\cite{jendal2023ginrec,wu2022inmo} and transformer-based~\cite{reimers2019sentencebert,shin2024attentivebsarec}, to \glspl{rs} based on variational encoders~\cite{zhao2022improvingcvar}.
However, a large pool of methods, as shown in \autoref{table:methods}, can make inductive recommendations for either only new users or only new items.
Hence, a method able to recommend to new users would still need to train on the full set of items and vice versa.
Meta-learning methods can recommend to new users and with new items bu shortly training on the new data~\cite{lee2019melu,finn17modelagmaml}.
Numerous techniques propose using subgraphs based on user-item pairs, alleviating the need for learned user and item embeddings; instead, using the graph structure and distances to generate embeddings~\cite{zhang2019inductiveigmc,zhang2022geometricgimc}.
However, constructing distinct subgraphs for each pair is prohibitively time-consuming and space-consuming when ranking items~\cite{jendal2023ginrec,wu2022inmo}.
Several methods use user meta-data to improve recommendations~\cite{wang2021priviledgepgd,cai2023userihgnn}, but such data is often unavailable or limited to a small user subset~\cite{rodriguez24userhistory}.
Privacy and data constraints limit interest in these methods.



\input{tables/03_methods}

Graph-based approaches use \glspl{gnn} to perform aggregation over all nodes in the graph.
To reduce the training overhead, GraphSAGE~\cite{hamilton2017inductive} applies node sampling during batch constructions, fixing the memory overhead. 
GraphSAGE was designed for node classification and thus does not support recommendation lists.
Among inductive recommender systems, INMO~\cite{wu2022inmo} 
instead learns initial embeddings for a subset of users and items, which all nodes must aggregate from for their representation.
Thus, it does not use node features.
Yet, for very large graphs, using only neighbor sampling was insufficient, and PinSAGE~\cite{ying2018graphpinsage} thus applied both sampling of the training graph and introduced a MapReduce framework to scale-out the computation.
Notably, PinSAGE is designed to recommend pins to boards, which can be translated to users and items; however, contrary to users, the boards are not explicitly modeled by PinSAGE and the method thus focuses on item-item recommendation exploiting in this way the collaborative signal. 
Instead of relying only on the collaborative signal, GInRec~\cite{jendal2023ginrec} proposes using \gls{kg} information, applying relation-specific gates for aggregation, and simply representing users by their neighbors. 
When subsampling the graph, we naturally remove both users and items for which we are still interested in recommending. 
Methods unable to handle such scenarios are, therefore, not relevant.
Consequently, the relevant recommenders that we can examine are PinSAGE~\cite{ying2018graphpinsage}, INMO~\cite{wu2022inmo}, and GInRec~\cite{jendal2023ginrec}.


\changed{
\textbf{\emph{Training efficiency.}}
Multiple approaches exist for reducing the graph sizes other than sampling\cite{hashemi2024graphreductionsurvey}:
\begin{enumerate*}[label=(\roman*)]
    \item graph sparsification removes edges and/or nodes to reduce the computational cost while preserving performance.
    Using top-k nodes or edges has been used based on various scoring metrics, such as PageRank~\cite{page99pagerank} or through a parameterized method~\cite{jin2022graphcondensationgcond}.
    \item Graph coarsening merges nodes into supernodes either through reconstruction or other optimization strategies~\cite{hashemi2024graphreductionsurvey}. 
    The reconstruction can be either through spatial, by merging pairs with the least effect on the reconstruction error, or spectral methods, by comparing the eigenvalues or vectors.
    Alternatively, it is possible to learn supernodes, representing a cluster of the graph~\cite{huang2021scalingcoarsening}.
    \item Graph condensation, constructs a synthetic graph for which a method can be trained on with similar performance~\cite{hashemi2024graphreductionsurvey}. 
    They use gradient matching between a method learned on the original graph and the synthetic graph, distribution matching of the properties, or trajectory matching.
\end{enumerate*}
Sparsification and condensation methods are usually designed for node classification and almost always rely on labeled nodes~\cite{hashemi2024graphreductionsurvey,sun2024gcbench}.
Furthermore, the condensation methods can be both time and space intesive~\cite{xiao2024disentangleddisco,sun2024gcbench}.
}

Sampling of graphs has been used for approximate spectral clustering~\cite{tremblay2020spectralclusteringsampling}, for topology estimation~\cite{kurant2012topologyestimationsampling}, estimating graph characteristics~\cite{cem2013graphcharacteristicssampling}, and covariance estimation~\cite{chepuri2017covarianceestimationsampling}.
However, none of these study node embedding methods. 
Many  \gls{gnn} methods apply sampling during training, requiring recomputation at each batch. 
They can be grouped largely into node-wise, layer-wise, and subgraph-based methods~\cite{liu2022sampling}.
Nevertheless, the sampling is always performed on the \textit{full train graph}, repeatedly, which can be infeasible in practice.
Within hyperparameter optimization, multiple methods exist to decrease tuning time, with one branch focusing on dataset sampling~\cite{montanari22recommendationhpo,garcia19samplinghpo}.
However, after finding optimal parameters, the methods still require training on the full graph due to working with transductive methods. 
Other works use sampling to adaptively select negative sampling for faster training; however, they still require all positive samples~\cite{chen2023datasubsamplingctr}.
Instead, PinSAGE~\cite{ying2018graphpinsage} was shown to be able to  train on a random uniform sampling over the graph.
Specifically, sampling $20\%$ of all graph boards that, for their dataset, proved to have negligible impact on performance.
However, the final graph after sampling still contained multiple millions of nodes and their graph was not the usual bipartite or multi-partite graph.
Thus, the question about which sampling technique is more effective and what are the actual effects on different dataset size and domains remains open.
For example, random node sampling would sample sporadic nodes and create loosely connected graphs, which is less suitable for graph convolution methods.

Therefore, for the first time, we study different graph-based sampling techniques for state-of-the-art inductive methods. 
We choose to focus on well established sampling methods that ensure semi-coherent graph structures~\cite{leskovec06samplinglargegraphs}. 

%% file: tables/03_methods.tex
{
\setlength{\textfloatsep}{0.6cm}
\definecolor{myGreen}{RGB}{77, 175, 74}
\definecolor{amber}{rgb}{1.0, 0.75, 0.0}
\newcommand{\cding}[2]{\textcolor{#1}{\ding{#2}}}
\newcommand{\yes}{\cding{myGreen}{52}}
\newcommand{\no}{\cding{red}{55}}
\newcommand{\maybe}{\color{amber}(\ding{52})}
\newcommand{\heightcmd}[1]{\rule[#1]{0pt}{#1}}
\newcommand{\meluheight}{1em}
\newcommand{\mcheight}{2em}
\scriptsize

\setlength\tabcolsep{1.2pt}
\renewcommand\extrarowheight{1.5pt}
\begin{table}[!t]
\caption{Related recommendation methods, the Task they support among (C) Node Classification, (R) Ranking, (P) Rating Prediction, and (SR) Sequential Recommendation.}
\label{table:methods}
\vspace{-5pt}
\centering
\resizebox{\linewidth}{!}{%
\begin{tabular}{l|c|cc|RRRA{2cm}|A{2cm}}
 & \textbf{} & \multicolumn{2}{c|}{\textbf{Inductive}} & \multicolumn{3}{R}{\textbf{Metadata}} & \textbf{} \\
\textbf{Model} & \textbf{Task} & \textbf{User} & \textbf{Item} & \textbf{User} & \textbf{Item} & \textbf{External} & \textbf{Architecture} & \textbf{Main Limitation} \\ \hline
BERT4Rec~\cite{sun2019bert4rec} & SR & \yes & \no & \no & \no & \no & Transformer & \multirow{5}{2cm}{\centering Cannot recommend for new items}  \\
IDCF~\cite{wu2021towardsidcf} & P & \yes & \no & \no & \no & \no & Matrix factorization &  \\
ReBKC~\cite{hui2022personalizedrebkc} & P & \yes & \no & \no & \no & \yes & Multi-headed attention & \\
IGCCF~\cite{damico2023itemigccf} & R & \yes & \no & \no & \no & \no & GNN & \\
BSARec\cite{shin2024attentivebsarec} & SR & \yes & \no & \no & \no & \no & Transformer & \\ \hline
ICP~\cite{zhang2021inductiveicp} & R & \no & \yes & \no & \yes & \no & NN & \multirow{3}{2cm}{\centering Cannot recommend for new users} \\
GAR~\cite{chen2022generativegar} & R & \no & \yes & \no & \yes & \no & Adversarial learning & \\
CVAR~\cite{zhao2022improvingcvar} & R & \no & \yes & \yes & \yes & \no & Variational encoder & \\ \hline
MeLU~\cite{lee2019melu} & R & \maybe & \maybe & \yes & \yes & \no & Meta-learning & \multirow{2}{2cm}{\centering Requires retraining for each new user} \\ 
MetaKG~\cite{du2023metakg} & R & \maybe & \maybe & \no & \no & \yes & Meta-learning & \\ \hline 
IGMC~\cite{zhang2019inductiveigmc} & P & \yes & \yes & \no & \no & \no & Subgraph & \multirow{2}{2cm}{\centering %
User-item subgraph construction is cost-intensive
} \\
GIMC~\cite{zhang2022geometricgimc} & P & \yes & \yes & \no & \no & \no & Subgraph & \heightcmd{-1em} \\ \hline
PGD~\cite{wang2021priviledgepgd} & R & \yes & \yes & \yes & \yes & \no & Student/teacher model & \multirow{2}{2cm}{\centering \heightcmd{.5em} Requires user metadata} \\ 
IHGNN~\cite{cai2023userihgnn} & R & \yes & \yes & \yes & \yes & \yes & GNN & \\ \hline
GraphSAGE~\cite{hamilton2017inductive} & C & \maybe & \yes & \no & \yes & \no & GNN & {\centering Not made for recommendation}   \\\hline
PinSAGE~\cite{ying2018graphpinsage} & R & \maybe & \yes & \no & \yes & \no & GNN w/ attention & \multirow{3}{2cm}{\centering } \\
INMO~\cite{wu2022inmo} & R & \yes & \yes & \no & \no & \no & GCN & \\
GInRec~\cite{jendal2023ginrec} & R & \yes & \yes & \no & \yes & \yes & GNN w/ gates & \\
\end{tabular}
}
\vspace*{-1.7\baselineskip}
\end{table}
}

%% file: sections/05_methodology.tex
We detail here the sampling methods and the inductive recommender systems used. 
For the recommenders, we describe only the most important parts contributing to their performance. 

\subsection{Sampling methods}
We evaluate two standard graph sampling approaches described as the most scalable and effective for reducing the size of very large graphs and designed specifically for their ability to preserve structural properties of the graphs~\cite{leskovec06samplinglargegraphs}.
The sampling technique adopted by PinSAGE~\cite{ying2018graphpinsage}, and a simple baseline taking into account the temporal information on edges. 
We perform node sampling for all methods, producing an induced subgraph where all connecting edges among the sampled nodes are preserved. 
When sampling from the \gls{kg}, we limit the starting nodes to nodes for the \gls{cg}.

\textbf{\emph{\acrfull{ff}~\cite{leskovec05graphsovertimeforestfire}.}}
\glsunset{ff} 
\gls{ff} simulates a tree burning process, where nodes ignite neighbors based on probabilities.  
It uses two edge probabilities: forward $\probability_f$ and backward $\probability_b$.  
We test two edge-sampling methods: \gls{ff}B with a binomial mean of $(1-\probability)^{-1}$~\citep{leskovec05graphsovertimeforestfire}, and \gls{ff} with mean $n\probability$.  
The latter is greedier in selecting edges when encountering a hub, thus terminating earlier, but it produces very skewed distributions, as shown in the experimental section.
Given a random starting node, the method ignites both backward and forward-going edges; the new burning nodes can now also burn their neighbors, and thus, the forest fire continues.
If no new burning nodes exist, a new random start node is selected. 
We present the \gls{ff} algorithm in \autoref{algo:forest_fire} to illustrate the use of the sampled input nodes.

\input{algorithms/joined_algo}

\textbf{\emph{\acrfull{rw} and \acrfull{rj}~\cite{page99pagerank}.}}
\glsunset{rw}\glsunset{rj}
\gls{rw} randomly selects a starting node and performs random walks from it with a restart probability $\probability_c$; adding visited nodes to the frontier. 
If at each step of the walk no new nodes could be visited, a new node is picked as the starting node. 
\gls{rj} is a similar method that randomly jumps to a new node during the walk, with the same probability $\probability_c$. 

\textbf{\emph{\acrfull{ps}~\cite{ying2018graphpinsage}.}\glsunset{ps}}
When training PinSAGE~\cite{ying2018graphpinsage},  ``board'' sampling is proposed for training using a smaller graph. 
In this case, the graph is a bipartite graph between boards and pins, and when a board is sampled, itself and all its pins are added to the sample until some criteria is met.
We adapt this sampling method by simply sampling users and their interactions for the \gls{cg}.
However, this only works for bipartite graphs, and adapting it to the \gls{kg} is non-trivial.
We use \gls{rw} for \gls{kg} sampling since a taxonomy path describes meaningful connections.
Further development for heterogeneous graphs remains an open research question.


\input{tables/07_dataset_n_kg_stats}

\textbf{\emph{\acrfull{ts}.}\glsunset{ts}}
As each rating is associated with a time $\ctime$, we can sample the users and items that have been active most recently. 
Meaning, given a \gls{cg}, we sample the user $\user$ and item $\rec$, s.t. the rating time is newer than that of any other user $\otheruser$ and item $\otherrec$ rating, as $\func_\ctime(\user,\rec)\geq\func_\ctime(\otheruser,\otherrec)$. 
Then, given that the \gls{kg} does not have any timestamps, we use \gls{rw} for that portion of the graph.

\changed{\textbf{\emph{Time Complexity.}}
The complexity of \gls{ff} is $O(|\verts{}|+|\relations{}|)$ when sampling all edges as the algorithm when starting at the root in a tree structured graph, as the algorithm is equivalent to breath first search. 
For \gls{rw} and \gls{rj}, the worst case would be a graph of disconnected nodes containing only self loops, $O(|\verts{}|lw)$, where $l$ is the walk length and $w$ is the number of walks performed per node. 
\gls{ps} goes through all users and their interactions, the complexity is thus $O(|\users|+|\relations{}|)$. 
Finally, for \gls{ts}, the complexity is $O(|\relations{}|)$ as the method in the worst case need to visit each edge.}

\subsection{Inductive recommenders}
The best-performing methods for recommendation in this setting are all based on \glspl{gnn} and perform graph convolutions.
A \gls{gnn} can be described using an aggregation function and an update function, the former computing a neighborhood representation of nodes and the latter updating the node~\cite{hamilton2017inductive}. 
For example, the neighborhood aggregation can be the mean of its neighborhood, followed by a non-linear layer as an update function~\cite{hamilton2017inductive}:
\begin{equation}\label{eq:general}
    \lemb{e}{\neighborhood{\cvert}}{l} = 
        \frac{1}{|\neighborhood{\cvert}|} 
        \sum_{(\othervert, \cvert)\in \neighborhood{\cvert}} 
        \lemb{e}{\othervert}{l-1}, 
    \lemb{e}{\cvert}{l} = \sigma\left(\cmatrix{w}\left[\lemb{e}{\cvert}{l-1}\|\lemb{e}{\neighborhood{\cvert}}{l}\right]\right),
\end{equation}
\noindent where $l\in [1, \ldots, L]$ is the current layer, $\neighborhood{\cvert}$ is the neighborhood of $\cvert\in\verts{}$, $\lemb{e}{\cvert}{l}\in \real^{d}$ is the embedding at layer $l$, $\sigma$ is some activation function, $\cmatrix{w}\in\real^{d'\times d}$ is a linear layer, and $[. \| .]$ is concatenation. 
The initial embeddings of $\lemb{e}{\cvert}{0}$ can thus be represented either by using a learned embedding or by extracting features. 
We refer to the initial embedding of all vertices as $\iembedding\in\real^{|\verts{}|\times d^0}$.
A \gls{gnn} captures information from distant nodes through multiple graph convolutions.  
Each convolution acts like a bounded BFS, so graph size directly impacts training time.  


\textbf{\emph{Pin SAmpling and aggreGatE (PinSAGE~\cite{hamilton2017inductive}).}}
PinSAGE's node embeddings are created, based on the GraphSAGE architecture, using the feature function $\feature$, allowing the method to perform inductive recommendation.
PinSAGE was used to recommend pins to boards, i.e.,  user collections of similar items.
Thus, the method only uses an item graph $\graph{\rec}$, where edges represent the co-pinning (co-interactions) of the items, thus not representing the users. 
For example in \autoref{fig:sampling_example}, ``The Prestige'' and ``Don Jon'' would be connected as Aiden likes both of them. 
Hence, PinSAGE optimizes towards item similarity as:
\begin{equation}
    \sum_{(\rec,\otherrec)\in\graph{\rec}}
    \mathbb{E}_{\otherrec' \sim Pr(\rec)}
    \text{max}(0, \hidden{\rec} \hidden{\otherrec'}- \hidden{\rec}\hidden{\otherrec}+\Delta),
\end{equation}
\noindent where $\Pr(\rec)$ is a probability of selecting a negative item given $\rec$. 

\textbf{\emph{Gated Inductive Recommender (GInRec~\cite{jendal2023ginrec}).}}
GInRec proposes using \gls{kg} information in addition to user interactions. 
The method uses relation-specific gates to capture the relational information.
GInRec further applies an auto-encoder architecture over all input features given by $\feature$ to reduce their dimensionality.
In contrast to PinSAGE, users are represented in the graph using the \gls{ckg} and initialized using a zero-vector, assuming graph convolutions are sufficient for user representations. 
The method can thus utilize \gls{bpr} loss for ranking~\cite{rendle2012bpr}, trying to rank positive items higher than negative items, which is optimized in conjunction with the auto-encoder loss.

\textbf{\emph{Inductive Module for collaborative filtering (INMO~\cite{wu2022inmo}).}}
INMO uses a key-query architecture, selecting a subset of nodes as keys, learning their representations and how to infer representations for non-key elements.
Hence, it defines a subset $\users_k{\subseteq}\subgraph{\users}$ and $\recs_k{\subseteq}\subgraph{\recs}$, that can be used to represent all users and items as:
\begin{align}
    \lemb{e}{\user}{0} &= 
        \frac{1}{\left( |\recs_\user \cap \subgraph{\recs}| + 1 \right)^\alpha} 
        \sum_{i\in \recs_u \cap \recs_k}\lemb{e}{\rec}{-1} + \cvector{e}_{user}, \\
    \lemb{e}{\rec}{0} &= 
        \frac{1}{\left( |\users_\rec\cap \subgraph{\users}| + 1 \right)^\alpha} 
        \sum_{\user\in \users_i\cap \users_k} \lemb{e}{\user}{-1} + \cvector{e}_{item},
\end{align}
\noindent where $\lemb{e}{\rec}{-1}\in\real^{d^{-1}}$ are the learned embeddings, $e_{user}$ is a learned template embedding and $\recs_\user = \{\rec|(\rec,\user)\in\neighborhood{\user}\}$. 
For item queries, the same equations are used, although they are inverted. 
Since all vertices in the \gls{cg} are represented as the average embedding, the individuality of the learned embeddings is lost.
Therefore, in tandem with the BPR loss, INMO proposes a self-enhancing loss:
\begin{equation}
    \sum_{\user\in\subgraph{\users}}
    \sum_{\rec\in\recs_\user\cap\subgraph{\recs}}
    \sum_{\otherrec\in \subgraph{\recs}\setminus\recs_\user} 
    ln \;\sigma\left( 
        \lembtop{e}{\user}{-1}\cmatrix{W}_s\lemb{e}{\rec}{-1} - 
        \lembtop{e}{\user}{-1}\cmatrix{W}_s\lemb{e}{\otherrec}{-1}
    \right)
\end{equation}
\noindent While user $\lemb{e}{\user}{0}$ and item $\lemb{e}{\rec}{0}$ embeddings can be used by any subsequent recommender, INMO adopted LightGCN~\cite{he2020lightgcn}.

%% file: algorithms/joined_algo.tex
{
\footnotesize
\begin{figure}[t]
\vspace{-1em}
\begin{minipage}{1\linewidth}
\begin{algorithm}[H]
    \caption{General sampling architecture}\label{algo:sampling}
    \begin{algorithmic}
        \Input{$\graph{cg}, \graph{kg}, \ratio, \textsc{Sampler}$}
        \Output{$\graph{cg}', \graph{kg}'$, where $|\relations{cg}|\cdot\ratio\approx|\relations{cg}'|\wedge|\relations{kg}|\cdot\ratio\approx|\relations{kg}'|$}
        \State $\graph{cg}' \gets \Call{Sampler}{\graph{cg}, \{\}, \ratio}$
        \State $\graph{kg}' \gets \Call{Sampler}{\graph{kg}, \verts{kg} \cap \verts{cg}'}$
    \end{algorithmic}
\end{algorithm}%
\end{minipage}

\vspace{-1em}

\begin{minipage}{1\linewidth}
\begin{algorithm}[H]
    \caption{Forest Fire algorithm}\label{algo:forest_fire}
    \begin{algorithmic}[1]
        \Require \textsc{SampleNeighbors}: samples neighbors of a node given probabilities and \textsc{NodeSubgraph}: constructs a subgraph containing only input nodes and the edges of the resulting graph. 
        \Function{ForestFire}{$\graph{}, \verts{in}, \ratio, \probability_f, \probability_b$}
            \State $e \gets |\graph{}| \cdot \ratio$ \Comment{Number of edges to sample} 
            \LineComment{Initialize burning, frontier, and \#samples}
            \State $B \gets \{\}, F \gets \{\}, s \gets 0$   
            \State $w \gets$ Random start node from ($F$ \textbf{if} $F\neq \varnothing$ \textbf{else} $\verts{}$)
            \While{$s \leq e \wedge s \leq |\graph{}|$}
                \State $N \gets$ \Call{SampleNeighbors}{$\graph{}, w, \probability_f, \probability_b$}
                \State $B \gets B \cup N \cup \{w\}$
                \State $F \gets F \cup \{w\}$
                \If{$\verts{in} \setminus F = \varnothing$}
                    \State $S \gets \verts{} \setminus F$ \textbf{if} $B \setminus F = \varnothing$ \textbf{else} $B \setminus F$
                    \State $w \gets$ Random node from $S$
                \Else
                    \State $w \gets$ Random node from $(\verts{in} \cup B) \setminus F$
                \EndIf
                \State $s\gets |\Call{NodeSubgraph}{\graph{}, B}|$
            \EndWhile
            \State \textbf{return} \Call{NodeSubgraph}{$\graph{}, B$}
        \EndFunction
    \end{algorithmic}
\end{algorithm}
\end{minipage}
\vspace{5pt}
\end{figure}
}

%% file: tables/07_dataset_n_kg_stats.tex
{
\footnotesize
\setlength\tabcolsep{1.2pt}
\begin{table*}[!th]
    \caption{Dataset properties: 
    I: items; 
    U: users; 
    R: ratings;
    TR: ratings in test set;
    STime: is the skewness of the rating times using the Fisher-Pearson coefficient;
    and DCG/DKG are the densities for the \acrlong{cg} and \acrlong{kg}, respectively.
    }
    \vspace{-5pt}
    \label{tab:datasets}%
    \centering
    \begin{tabular}{l|rrrrrrRRRRRRRZ}
     & \textbf{\#I} & \textbf{\#U} & \textbf{\#R} & \textbf{DCG} & \textbf{\#TR} & \textbf{STime} & \textbf{OLO} & \textbf{NLO} & \textbf{\#OTU} & \textbf{\#OTI} & \textbf{\#NTU} & \textbf{\#NTI} & \textbf{\#NTI NR} \\
    \hline
    \textbf{MovieLens} & 4,645 & 14,206 & 1,889,382 & 2.86E-02 & 499,040 & 0.26 & 5.61 & 110.38 & 13,377 & 4,379 & 829 & 51 & 19 \\
    \textbf{Amazon Book} & 24,841 & 70,679 & 843,228 & 4.80E-04 & 322,048 & -0.90 & 1.02 & 4.06 & 67,800 & 11,847 & 2,867 & 210 & 99 \\
    \textbf{Yelp} & 77,319 & 174,840 & 2,428,509 & 1.80E-04 & 809,989 & -0.50 & 1.13 & 5.77 & 168,144 & 40,420 & 6,696 & 1,668 & 1,173 \\
\end{tabular}%
\begin{tabular}{Zrrrrr}
     & \textbf{\#Entities} & \textbf{\#Relations} & \textbf{\#Relationships} & \textbf{DKG} \\
    \hline
    \textbf{MovieLens} & 14,062 & 8 & 100,719 & 2.88E-04 \\
    \textbf{Amazon Book} & 88,572 & 39 & 2,555,995 & 1.99E-04 \\
    \textbf{Yelp} & 75,199 & 12 & 1,643,792 & 7.07E-05 \\
\end{tabular}
    \vspace{-5pt}
\end{table*}
}

%% file: sections/06_experiments.tex
As in PinSAGE~\cite{ying2018graphpinsage}, we aim at reducing the amount of resources needed and the computation cost of training by using a subsampled graph, while inference is still performed on the full graph.
We answer the following questions:
\begin{enumerate*}[label=\textbf{RQ\arabic*)}]
    \item How do sampling methods affect the ability of \glspl{rs} to learn reliable models?
    \item How does sample size affect the models' performance?
    \item What is the correlation between training time and performance when sampling?
    \item How do different \glspl{rs} models handle subsampling?
\end{enumerate*}

\subsubsection*{Datasets}
We evaluate the methods on three real-world datasets (See~\autoref{tab:datasets}):
\begin{enumerate*}[label=(\roman*)]
    \item a dataset with ratings on movies, \gls{ml}~\cite{harper2015movielens} extended with the MindReader \gls{kg}~\cite{brams2020mindreader};
    \item one with reviews of books, \gls{ab}~\cite{ni2019justifyingamazonreviewdataset} for which a \gls{kg} was constructed when testing the transductive method KGAT~\cite{wang2019kgat}; and
    \item a dataset with reviews of businesses, \gls{yd}~\cite{yelpdataset}, for which we extracted a \gls{kg}~\cite{corfixen2023yelpknowledgegraph}.
\end{enumerate*}
For each dataset, we only use ratings for items connected to the respective \glspl{kg}, removing all other items and the respective ratings.
We further remove users and items with less than 5 ratings as well as users with ratings spanning less than 5 days.
The datasets are split with ratios $0.8{:}0.1{:}0.1$ for train, validation, and testing, respectively; \changed{ensuring that all ratings of the train set occur before the validation set, and validation before test.
This ensures trends occur naturally over time and that all methods are tested on new data.}
The sampling is performed on the training partition, as we are interested in reducing the training time.
We sample a few ratings for each user for validation and testing, simulating new users being greeted with an initial page where they provide initial ratings, similarly to~\cite{jendal2023ginrec,lee2019melu}.

Analyzing the datasets (\autoref{tab:datasets}), we notice that the ratings are unevenly distributed over time. 
We report their skewness in the STime column. 
Positive values mean most ratings occur early, with less activity later; negative values indicate the inverse. 
We observe that most of the ratings for \gls{ab} occur late, while most occur early for \gls{ml}.
Such distributions naturally affect the subsequent results, where \gls{ab} and \gls{yd} probably adhere to normal business growth.
However, the \gls{yd} was affected by COVID, as seen in \autoref{fig:dataset_popularity}, tracking the number of ratings given to an item per month, with a moving window of 12 months.
Making the dataset non-trivial for \glspl{rs}.

\input{tables/05_sampling_statistics}
\input{figures/06_combined_plot}

\subsubsection*{Parameters}
For the samplers, we study different ratios $\ratio {\in} \{0.05, 0.1,\\0.2, 0.5, 1\}$, going from an extreme sub-sampling setting to the full graph. 
\changed{ 
Due to space constraints, full details for $0.05$ and $0.1$ are reported in the extended version on the online repository, while their implications are still discussed below.
}
As in related works~\cite{leskovec06samplinglargegraphs,leskovec05graphsovertimeforestfire}, we set the forward probability at $\probability_f=0.35$, the backward probability at $\probability_b=0.2$, the jump/restart probability at $\probability_c=0.15$, and set the walk length at $10$.
We have implemented all methods in PyTorch,
testing each method's implementation until we achieved a similar performance on the original datasets reported. 
For parameter tuning, we apply \gls{asha}~\cite{li2020massivelyasha}, a method based on multi-armed bandit methodology of high initial exploration of parameter combinations, before focusing on fewer combinations.
Methods are tuned on the full graphs, using the same hyperparameters in all sampling settings. 
\changed{We tried tuning on $\ratio\in\{0.2,0.5\}$ using \gls{ps} getting similar performance as tuning on the full graph on these ratios. 
We studied \gls{ps} as it has been used in the industry and want to validate its performance in other settings~\cite{ying2018graphpinsage}.
All configurations and additional experimental results are available in the online repository.}
We use an NVIDIA A10 GPU, dual processor setup with Intel Xeon Gold 6326, and 256 GB RAM.

\subsubsection*{Features}
For feature extraction, we utilize the average textual embeddings of Sentence-BERT~\cite{reimers2019sentencebert} for all items, with the texts being the first paragraph of the Wikipedia page, when available, otherwise Wikidata, for \gls{ml} and \gls{ab}, and review text for \gls{yd}.
Furthermore, we compute the node degrees normalized by centering around zero and scaling to unit variance. 
The scaling is calculated for the train features and applied to the validation and test features. 
However, the descriptions of entities can be non-descriptive (see \url{https://www.wikidata.org/wiki/Q20656232}).
We, therefore, use TransR~\cite{lin2015learningtransr} to generate embeddings for all descriptive entities.

\subsubsection*{Evaluation metrics}
We rank all items in the test set, as ranking a subset has been shown to skew the results~\cite{rendle2019evaluationmetrics}.
We exclude items already interacted with since multiple ratings between the same user and item cannot occur~\cite{wang2019kgat}.
We use four standard ranking measures: NDCG@k, recall@k, precision@k, and PR-AUC, and one serendipity measure, coverage@k~\cite{adomavicius2012improvingdiversitycoverage}; reporting the average performance over all users.
A high coverage is not indicative of the recommendation performance, as a \gls{rs} giving random recommendations would have high coverage but low ranking ability; therefore, it cannot be looked at in isolation.
However, for brevity, \emph{we only report HR and NDCG} in the comparison table~\autoref{tab:results}, as the other results confirm the same findings we report here.

\subsubsection*{Sampling viability} 
We report the sampling time and maximal memory usage during a run (implemented in Python without any parallelization) for sampling ratio $0.5$ in \autoref{tab:sample_stats}.
We note that it is possible to greatly optimize the sampling process for even faster running times.
Yet, we see that their running time is already orders of magnitude shorter than the respective training time, confirming the possible gain in runtime reduction when adopting sampling.

\changed{We see that \gls{ff}B is often the method with the longest running time.
That is because its sampling strategy selects only few nodes at every iteration, i.e., each step burns only one or two neighbors, meaning a higher number of iterations compared to a sampler constructing a densely connected neighborhood.}
To select more nodes we tried increasing the sampling probabilities s.t. the binomial mean would be around 10.
With higher sampling likelihood, the sampling speed reduces to that of \gls{rj}.
Furthermore, we see that the \gls{kg} sampling of \gls{ff} is also considerably faster than that of other methods for the \gls{ab} dataset.
This is due to presence of nodes with high degree.
We analyzed the degree distribution and found a great skew in the degrees, as the $99.99th$ percentile has around $~2k$ edges, while the $99.999th$ percentile has $~100k$ edges in the \gls{ab} \gls{kg}. 
We note sampling large portions of edges of few hubs is not desirable leading to poor distribution similarities as seen for \gls{ab} in \autoref{fig:dstatistics}.

We use the Kolmogorov-Smirnov D-statistic over the \textit{cumulative distribution function} of the in- and out-degrees for comparing the shapes of the distributions to evaluate the sampling methods ability to maintain representative structural properties of the original graph~\cite{leskovec06samplinglargegraphs}.
Lower values indicate a greater alignment between the sampled graph and the original graph.
We sample for each sampling method and ratio combination five times, plotting the D-statistics for each node-type in~\autoref{fig:dstatistics}.
Due to the heterogeneous nature of the \gls{kg}, the non-uniform degree distributions, and the restriction to consider already samples entities, the sampling methods have more difficulty in approximating the structural properties of the KG. 
Nonetheless, it would not make sense to relax the restriction on the seeding of the sample for the \gls{kg}, since information detached from the items would be irrelevant for item recommendation.
Current sampling methods treat edges and nodes as homogeneous.  
They ignore node types, likely causing distribution misalignment across types.  
Future work should develop methods tailored to bipartite \glspl{cg} and heterogeneous graphs like \glspl{kg}.

\subsubsection*{RQ1 \& RQ2} 
The summary of the result in terms of  NDCG and AUC compared to the reduction in training time is reported in \autoref{tab:results}.
We observe similar trends also for the other metrics not reported here.
\changed{
\textbf{Choice of the best sampler depends on the dataset, ratio, and \gls{rs} used.}
}
Interestingly, while \gls{ps} is able to better capture the \gls{cg}'s degree distribution (\autoref{fig:dstatistics}), \gls{ts} is the best performing of the two in most cases. 
Further, we notice that both \textbf{\gls{ps} and \gls{ts} perform well in many settings}, often being best or second best performing.
For example, for the \gls{ab} dataset, \gls{ts} and \gls{ps} are the best performing for all recommender methods for most sampling ratios. 
Yet, we see a trend where \gls{ts} becomes the best performing for all datasets and methods when $\ratio\leq 0.1$. 
This indicates that \textbf{for small datasets data recency is important} compared to capturing correct distributions and that it is very important in sparse situations for \gls{cf}. 

As it would be expected, in most cases, reducing the amount of training data reduces the prediction quality of the models.
This is most clearly seen on the \gls{yd}, where all methods have trouble generalizing properly without the full datasets, regardless of the sampling method.
When sampling $50\%$ of the graph, the final prediction qualities across the sampling methods present only limited differences. 
For \textbf{high ratios the sampling method is less important} as sufficient users, items and ratings have been sampled regardless of sampling methodology.
Yet, as previously stated, this is not true at lower ratios. 
This raises the question on the data-efficiency of these methods, i.e., whether these recommendation systems are actually able to infer inductive bias from complex graph structures or are just aggregators of collaborative signal (discussed in RQ4).
When \textbf{sampling $\mathbf{5\%}$ to $\mathbf{10\%}$, none of the methods can recommend well compared to their baseline performance on the whole graph}. 
The only exception is for the dense and popularity biased \gls{ml} dataset. 
\changed{
However, all methods perform better than a na\"ive TopPop recommender~\cite{cremonesi2010topnperformance} with only 5\% of the data, except GInRec on \gls{yd}.
}
Only INMO is capable of decent recommendations when using $20\%$ on the \gls{ab} dataset and maintains performance with $50\%$ of the data for the \gls{yd}.
\changed{We find \textbf{neither node-type distribution nor rating time similarity correlate with ranking performance}, as illustrated in \autoref{fig:ratioskew}. 
For example, for \gls{yd} with $r=0.1$, we find that \gls{ps} performs better than \gls{rw} in all cases; however, \gls{rw}'s user ratio is far closer to the train graphs ratio than \gls{ps}'s. 
For rating distribution, there is little difference between \gls{ts} and \gls{ps} performance.
Yet, \gls{ps} almost perfectly matches the base graphs' rating skew while \gls{ts} is far off.
There appears to exist a \textbf{ slight correlation between the degree distribution \gls{cg} and the performance of the sampler} exists, as \gls{ps} and \gls{ts} are frequently best performing and have the lowest d-statistics.
However, the absolute value is not indicative of performance and cannot be compared across datasets. 
}

\input{tables/04_results_rel}
\input{tables/08_actionable_insights}

\subsubsection*{RQ3} 
Generally, reducing the amount of training data significantly reduces the training time.
Yet, when using $50\%$ of the data, PinSAGE's training time does not decrease at the same rate of other methods.
This is likely due to the item-item loss function.
For \textbf{PinSAGE and GInRec we see that longer running times often correspond to better ranking performance}.
These methods jointly learn input feature representations, aggregation functions, and ranking composition.  
Therefore, learning to represent and aggregate node features may require more computation and data to learn inductive biases. 
In contrast, \textsc{INMO} only optimizes initial node embeddings for the ranking objective.   
The reduced training time is particularly important for INMO, which does not use node sampling, having exponential training time w.r.t. the number of edges (see \autoref{fig:resultsplot}).
INMO has a time complexity of $O(L|\relations{}|d)$ for propagation.
This is required for each edge in the train graph, leading to an epoch complexity of $O(L|\relations{}|^2d)$, where $L$ is the number of layers and $d$ the dimensionality.

On \gls{yd}, we observe a $4\%$ decrease in performance and $~86\%$ decrease in training time.
Although it seems like a substantial decline, \textbf{a $4\%$ decrease may effectively be negligible in practice}.
As the HR for INMO is $0.17$ and decreases to $0.163$ when $r{=}0.5$, this means that in a ranked list of 20 items, the method would show at least one relevant item in both settings (on average). 
Even when INMO's HR performance is reduced by $30\%$ for $r{=}0.05$, it still recommends at least one relevant item in the top 20, with a HR of $0.119$.
Furthermore, the sampling maintains the ordering of methods in almost all cases.
Thus, it is \textbf{generally possible to compare models on the sample} to infer their relative performance on the full graph~\cite{garcia19samplinghpo,montanari22recommendationhpo}. 
\changed{  
    In almost all cases, the methods performance with $r\in\{0.05,0.10\}$ was better than a na\"ive TopPop~\cite{cremonesi2010topnperformance} recommender, only GInRec performing worse on YD. 
    \textbf{INMO obtained up to 5x the performance for NDCG on $0.05$ compared to TopPop.}
}

Finally, we find \glspl{rs} give \textbf{less diverse recommendations when trained with less data}.
In \autoref{fig:resultsplot}, we highlight how INMO's coverage increases as more data is given to the method (similar trends are exhibited by the other baselines).
More training data allow the method methods' to make diverse recommendations 
A small subset of items may be sufficient for the \gls{ml} dataset, but has high impact on the performance for other datasets. 
Future research should \textbf{ensure \glspl{rs} are capable of providing diverse recommendations even when little data is available}.


\subsubsection*{RQ4} 
INMO seems to be the most robust method to downsampling.
Interestingly, while GInRec uses external knowledge, this seems to reduce the effectiveness of the method when there is little information available.
\textbf{The current methods cannot effectively exploit the \glspl{kg}' signal}.
Interestingly, PinSAGE performs best on \gls{ml}, likely due to \gls{ml} being popularity biased and as the textual attributes of movies are of far higher quality than what is available for both \gls{ab} and \gls{yd}.
However, all methods struggle on the \gls{yd}.
The dataset is likely more difficult compared to others as a dramatic change in most popular items occurs due to COVID, as seen in \autoref{fig:dataset_popularity}.
\changed{
To investigate whether methods prioritize collaborative signal or structural information, we design a Nich\'eSampler (NS) sampling items with the fewest user ratings.
For methods that learn inductive bias, the features present in the graph used for training dictates which inductive biases are learned by the model. 
Thus, with a training graph with only niche items, if the methods maintains the same predictive performance, it is because they rely only on collaborative signal, disregarding learned rules.
Looking at \autoref{tab:results}, we observe that all methods show a performance drop under NS, but GInRec and PinSAGE see a larger drop in performance.
Therefore, \textbf{all methods learn some form of inductive bias}.
Yet, INMO likely relies more on the collaborative signal, which is expected as it does not have any attention mechanisms on edges.
This means that \textbf{collaborative filtering can be learned from small samples}, while more robust inductive biases from graph structures requires different sampling techniques and learning architectures.
\gls{ts} and \gls{ps} perform well at lower ratios, suggesting they capture meaningful signals. 
Thus, leveraging time and representative user selection offers a promising path for improved sampling.

}

%% file: tables/05_sampling_statistics.tex
{
\footnotesize
\setlength\tabcolsep{1.2pt}

\begin{table}[!t]
        \caption{Sampling resource usage with sampling ratio $r=0.5$. The sampling time (Time) for CG/KG is in seconds, peak memory (Mem) in GB, and MTT is the floored average method train time. TC is time complexity with $l$ and $w$ being walk length and number of walks, respectively.}
\label{tab:sample_stats}%
\vspace{-5pt}
\begin{tabular}{l|rrr|rrr|rrr|c}
 & \multicolumn{3}{c|}{\textbf{ML}} & \multicolumn{3}{c|}{\textbf{AB}} & \multicolumn{3}{c|}{\textbf{YD}} & \multirow{3}{*}{\textbf{TC}}  \\
 & \multicolumn{2}{c}{\textbf{Time}} & \multicolumn{1}{c|}{\multirow{2}{*}{\textbf{Mem}}} & \multicolumn{2}{c}{\textbf{Time}} & \multicolumn{1}{c|}{\multirow{2}{*}{\textbf{Mem}}} & \multicolumn{2}{c}{\textbf{Time}} & \multicolumn{1}{c|}{\multirow{2}{*}{\textbf{Mem}}} &  \\
 & \multicolumn{1}{c}{\textbf{CG}} & \textbf{KG} & \multicolumn{1}{c|}{} & \multicolumn{1}{c}{\textbf{CG}} & \textbf{KG} & \multicolumn{1}{c|}{} & \multicolumn{1}{c}{\textbf{CG}} & \textbf{KG} & \multicolumn{1}{c|}{} &  \\ \hline
\textbf{FF} & <1s & \multicolumn{1}{r}{5s} & 0.9GB & 46s & \multicolumn{1}{r}{<1s} & 1.2GB & 192s & \multicolumn{1}{r}{80s} & 1.5GB & $|\verts{}|+|\relations{}|$ \\ \hline
\textbf{FFB} & 17s & \multicolumn{1}{r}{15s} & 0.9GB & 290s & \multicolumn{1}{r}{504s} & 1.2GB & 1,272s & \multicolumn{1}{r}{1,108s} & 1.5GB & $|\verts{}|+|\relations{}|$ \\ \hline
\textbf{PS} & 2s & - & 1.2GB & 12s & - & 1.2GB & 66s & - & 1.5GB & $|\users|+|\relations{}|$ \\ \hline
\textbf{RJ} & 8s & \multicolumn{1}{r}{10s} & 0.9GB & 96s & \multicolumn{1}{r}{194s} & 1.3GB & 442s & \multicolumn{1}{r}{298s} & 1.6GB & $|\verts{}|lw$ \\ \hline
\textbf{RW} & 8s & \multicolumn{1}{r}{11s} & 0.8GB & 58s & \multicolumn{1}{r}{100s} & 1.2GB & 152s & \multicolumn{1}{r}{109s} & 1.5GB & $|\verts{}|lw$ \\ \hline
\textbf{TS} & 4s & - & 0.9GB & 22s & - & 1.3GB & 74s & - & 1.4GB & $|\relations{}|$ \\ \hline
\textbf{MTT} & 3,860s & - & \multicolumn{1}{c|}{-} & 15,670s & - & \multicolumn{1}{c|}{-} & 77,728s & - & \multicolumn{1}{c|}{-} & 
\end{tabular}                                                                    
\vspace{-5pt}
\end{table}
}

%% file: figures/06_combined_plot.tex
\begin{figure*}
    \centering
    \begin{minipage}[b][][b]{.25\textwidth}
        \vspace*{\stretch{1}}
        \includegraphics[width=.92\linewidth,trim=0 10 0 0]{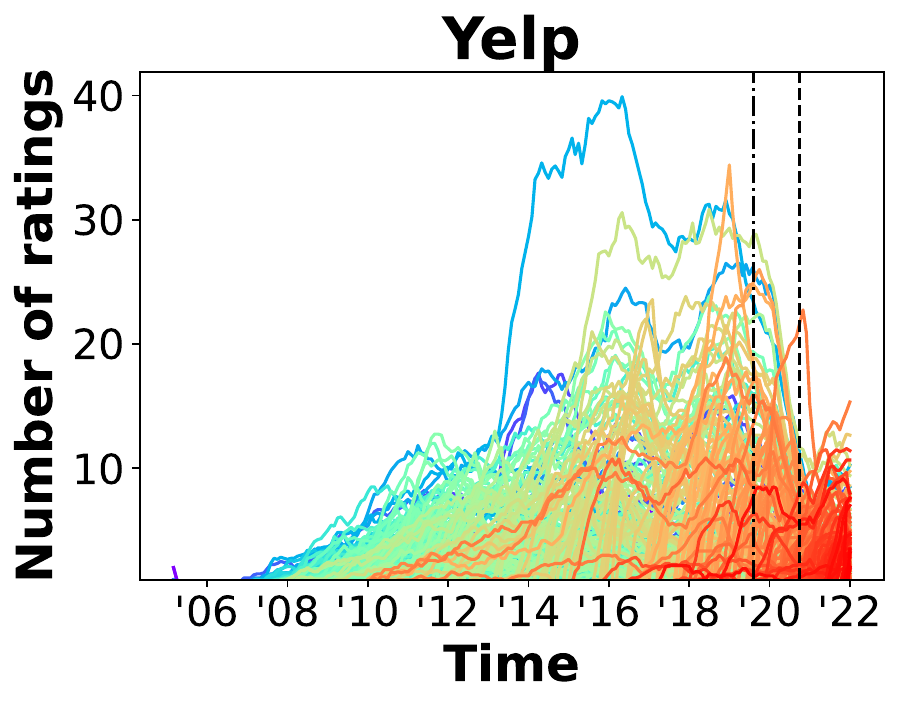}%
        \caption{\#Ratings on items.}%
        \label{fig:dataset_popularity}%
    \end{minipage}%
    \begin{minipage}[b][][b]{.75\textwidth}
        \centering
        \includegraphics[height=.33cm]{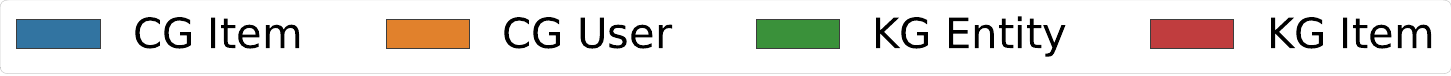}
        
        \settoheight{\imageheight}{\includegraphics[width=0.33\textwidth]{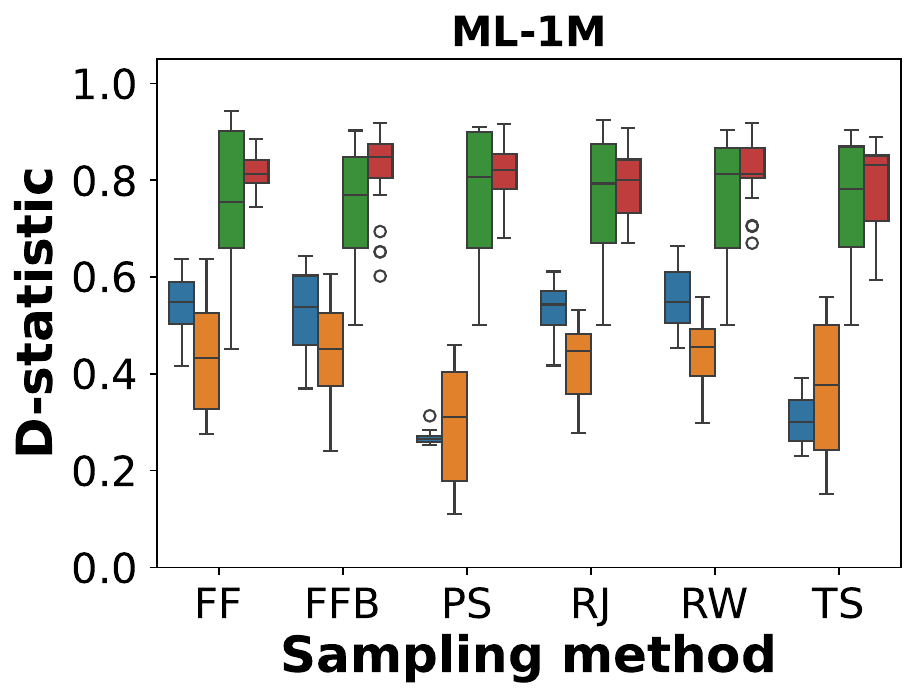}}%
        \includegraphics[width=0.33\textwidth]{figures/ML-1M_dstatistic.pdf}
        \includegraphics[height=\imageheight, trim=30 0 0 0, clip]{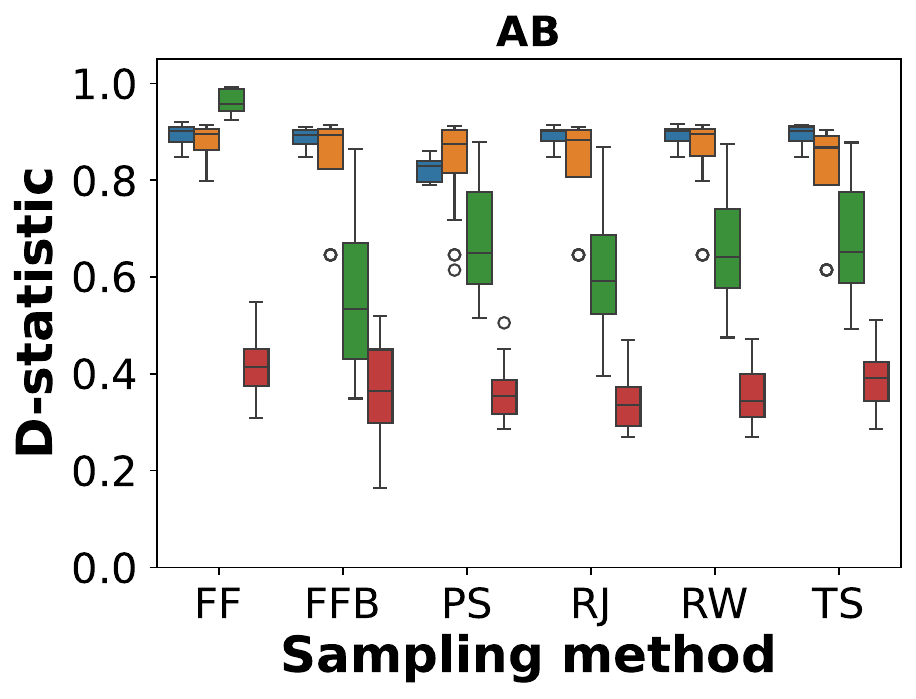}%
        \includegraphics[height=\imageheight, trim=30 0 0 0, clip]{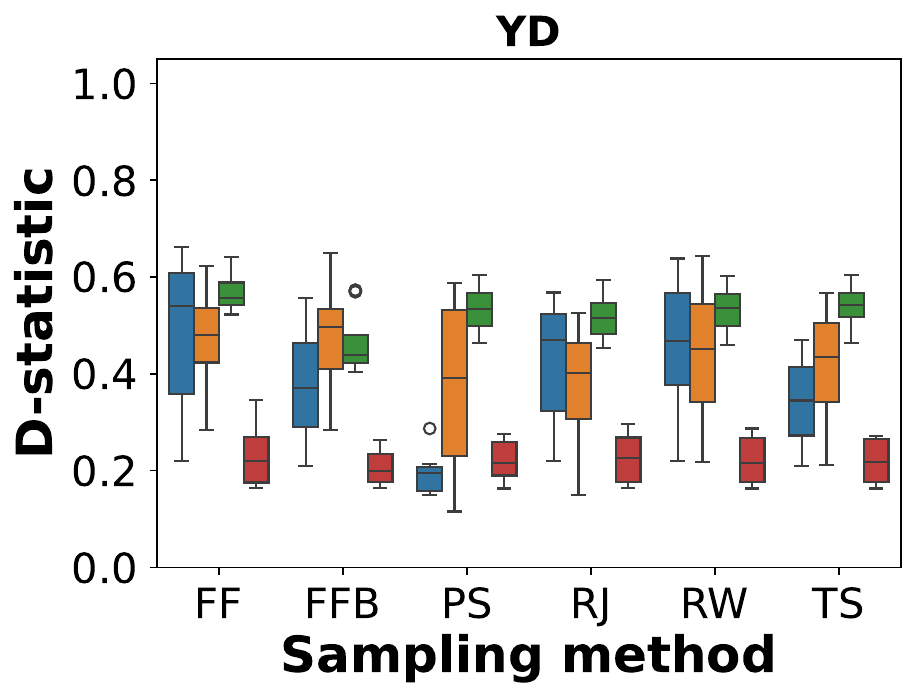}%
        \vspace{-5pt}
        \caption{\changed{Kolmogorov-Smirnov D-statistic of degree distribution per node type.}}
        \label{fig:dstatistics}
    \end{minipage}
    \begin{minipage}[b]{.5\textwidth}
        \centering
        \includegraphics[height=.33cm]{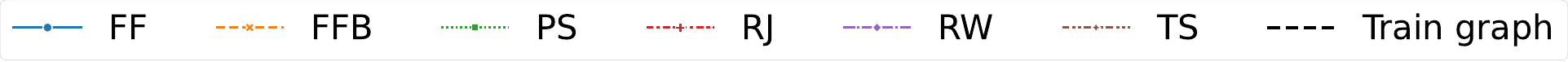}\\
    
        \settoheight{\imageheight}{\includegraphics[width=0.5\textwidth, trim=0 0 0 0]{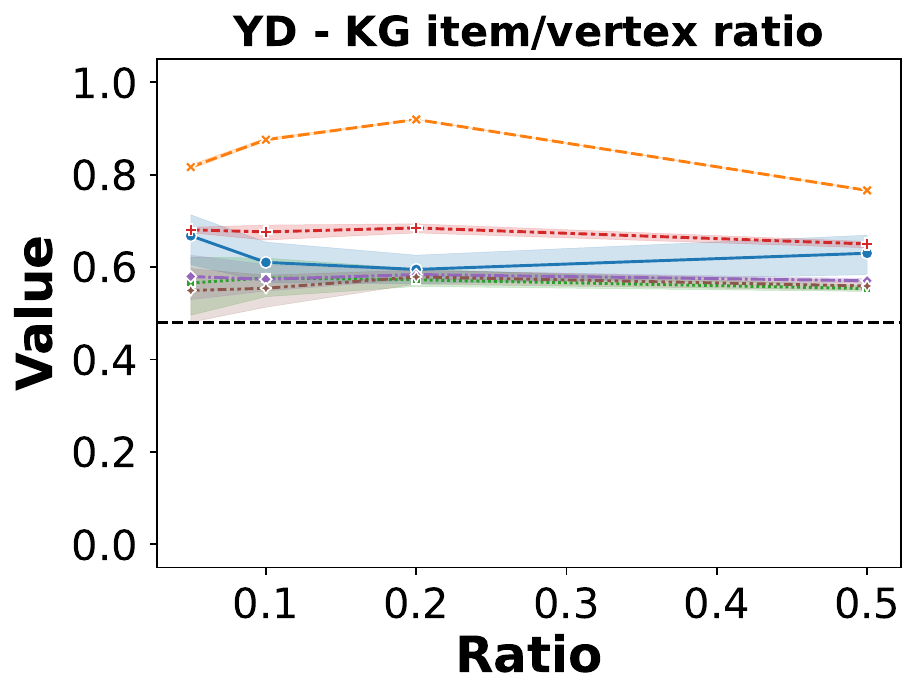}}

        \includegraphics[height=\imageheight, trim=30 0 0 0, clip]{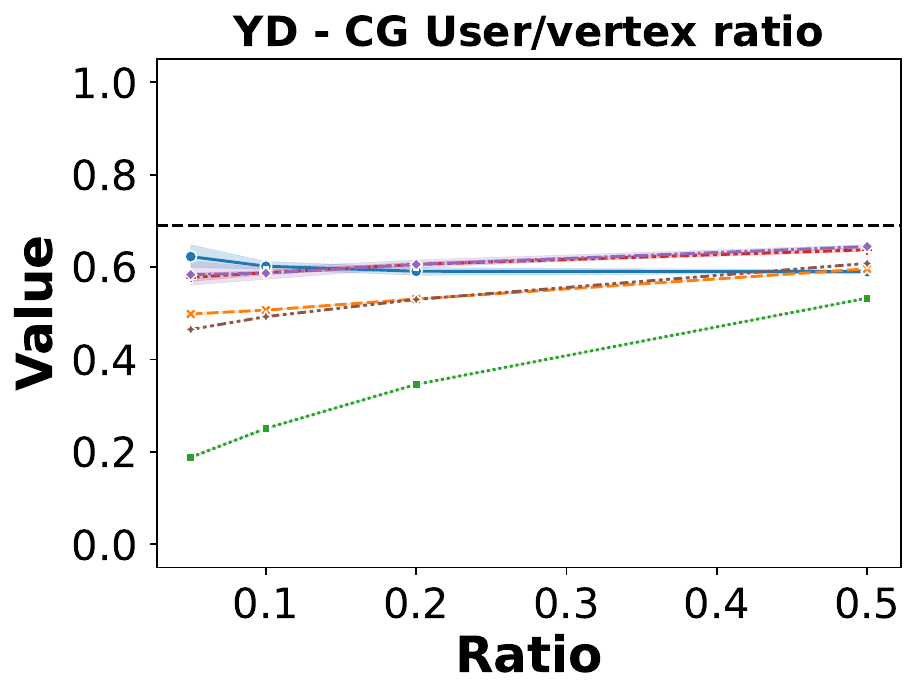}%
        \includegraphics[height=\imageheight, trim=30 0 0 0, clip]{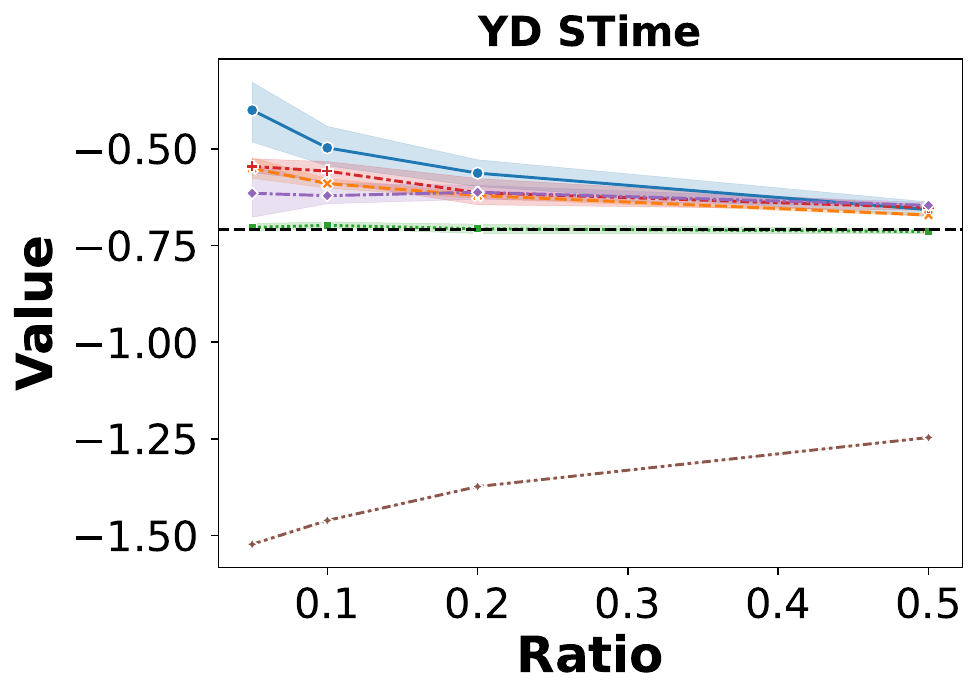}%
        \vspace{-5pt}
        \caption{\changed{Illustration of user ratio and rating skew in the YD.}}
        \label{fig:ratioskew}
    \end{minipage}%
    \begin{minipage}[b]{.5\linewidth}
        \centering
        \includegraphics[height=.33cm,clip=true,trim=60 0 590 0]{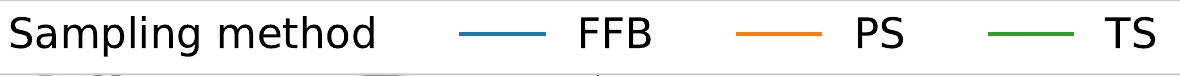}
        \includegraphics[height=.33cm,clip=true,trim=700 0 2 0]{figures/metric_errorbar_legend.pdf}\\
        \includegraphics[width=0.5\textwidth, trim=00 0 0 0, clip]{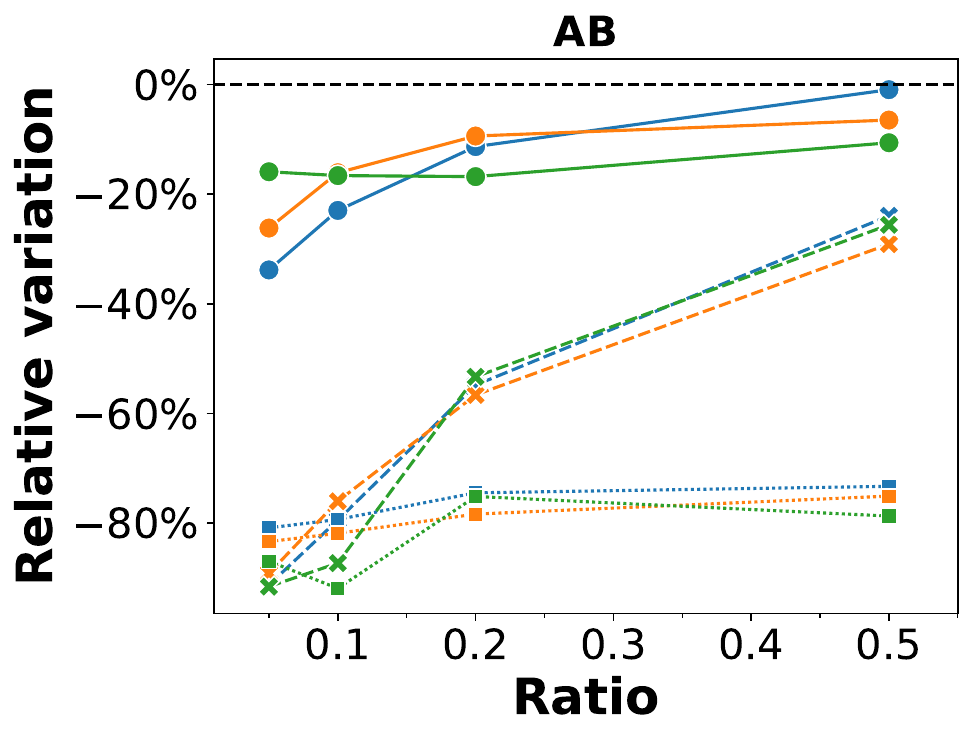}%
        \includegraphics[width=0.5\textwidth, trim=30 0 0 0, clip]{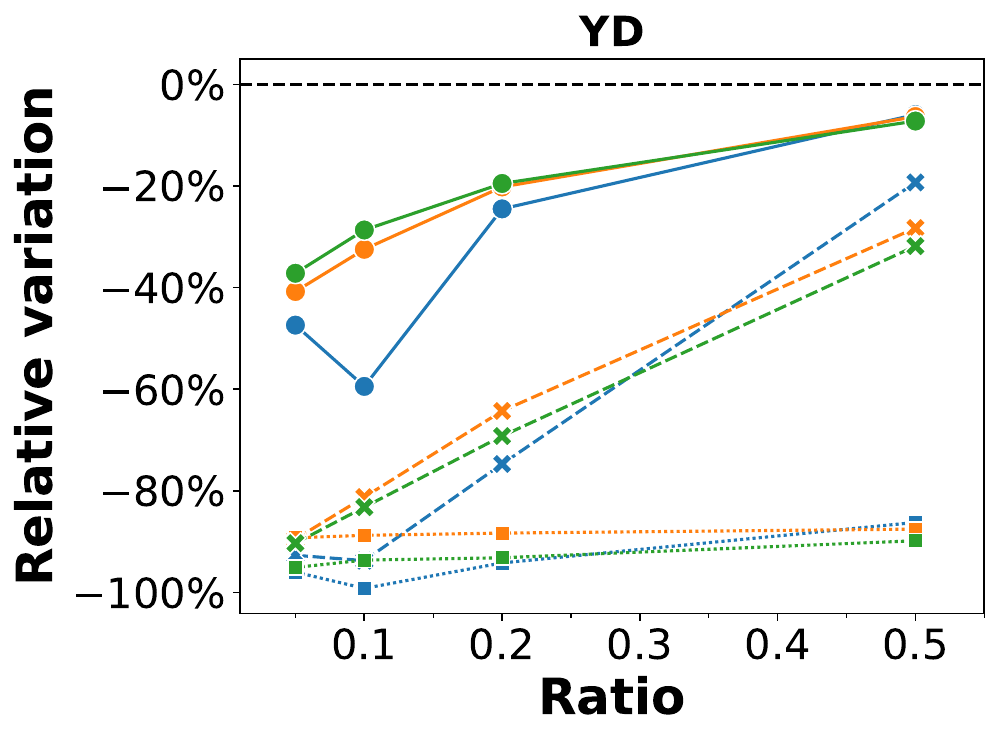}%
        \vspace{-5pt}
        \caption{INMO's relative performance.}
        \label{fig:resultsplot}
    \end{minipage}
    \vspace{-5pt}
\end{figure*}

%% file: tables/04_results_rel.tex
{
\setlength\tabcolsep{1.2pt}
\begin{table*}[!t]
    {\centering
    \caption{Results of methods at different sampling ratios and with different methods. All results are measures at $k=20$, except AUC which evaluates the complete list, with running time in hours. Bold indicates the best performing within a group.}
    \label{tab:results}
    \vspace{-5pt}
    \resizebox{1\linewidth}{!}{%
    \input{tables/results/sampling_rel_mrs_v2}
    }}
    {\raggedright \scriptsize\color{shade}\text{* Defined in RQ4.}} \hspace{50em}
    \vspace{-5pt}
\end{table*}
}

%% file: tables/results/sampling_rel_mrs_v2.tex
\begin{tabu}{Rrl|rrRRRRr|rrRRRRr|rrRRRRr|}%
 &  &  & \multicolumn{21}{c|}{\textbf{GInRec}} \\
 \cline{4-24}
 &  &  & \multicolumn{7}{c|}{\textbf{ML-1M}} & \multicolumn{7}{c|}{\textbf{AB}} & \multicolumn{7}{c|}{\textbf{YD}} \\
 &  &  & \textbf{HR} & \textbf{NDCG} & \textbf{Cov} & \textbf{Re} & \textbf{Pr} & \textbf{AUC} & \textbf{Time} & \textbf{HR} & \textbf{NDCG} & \textbf{Cov} & \textbf{Re} & \textbf{Pr} & \textbf{AUC} & \textbf{Time} & \textbf{HR} & \textbf{NDCG} & \textbf{Cov} & \textbf{Re} & \textbf{Pr} & \textbf{AUC} & \textbf{Time} \\
\hline
\multirow[c]{25}{*}{\textbf{GInRec}} & \textbf{-} & \textbf{-} 
& \textbf{0.734} & \textbf{0.202} & \textbf{0.355} & \textbf{0.103} & \textbf{0.168} & \textbf{0.111} & \textbf{0.9} 
& \textbf{0.132} & \textbf{0.029} & \textbf{0.390} & \textbf{0.059} & \textbf{0.008} & \textbf{0.013} & \textbf{2.7} 
& \textbf{0.122} & \textbf{0.021} & \textbf{0.135} & \textbf{0.036} & \textbf{0.008} & \textbf{0.009} & \textbf{7.9} \\
\cline{2-24}
\textbf{} & \multirow[c]{6}{*}{\textbf{0.20}} & \textbf{FF} & \underline{-5.1\%} & -10.3\% & -33.6\% & \underline{-12.1\%} & -7.9\% & -12.8\% & -75.5\% & -34.2\% & -33.9\% & -5.3\% & -39.2\% & -38.3\% & -28.0\% & -82.6\% & -62.5\% & -68.0\% & 94.8\% & -68.7\% & -64.7\% & -70.3\% & \textbf{-84.3\%} \\
\textbf{} & \textbf{} & \textbf{FFB} & -9.5\% & -12.6\% & -32.6\% & -18.5\% & -9.3\% & -14.1\% & -78.0\% & \underline{-31.0\%} & -30.1\% & \textbf{44.9\%} & \underline{-34.5\%} & \underline{-35.7\%} & -24.5\% & -79.5\% & \underline{-58.2\%} & \underline{-64.0\%} & \textbf{174.5\%} & \underline{-65.6\%} & \underline{-60.4\%} & \underline{-65.4\%} & -61.6\% \\
\textbf{} & \textbf{} & \textbf{PS} & -6.9\% & \underline{-8.7\%} & \underline{-31.5\%} & -14.5\% & \underline{-6.1\%} & \underline{-9.8\%} & -68.4\% & -38.8\% & -29.3\% & \underline{29.5\%} & -42.1\% & -45.2\% & \underline{-10.7\%} & \underline{-89.1\%} & -61.5\% & -68.9\% & 61.1\% & -69.1\% & -65.0\% & -70.2\% & \underline{-82.5\%} \\
\textbf{} & \textbf{} & \textbf{RJ} & -10.0\% & -15.7\% & -45.0\% & -18.7\% & -11.6\% & -17.5\% & \textbf{-83.3\%} & -35.8\% & -37.6\% & 20.5\% & -39.7\% & -40.7\% & -34.1\% & -81.6\% & -64.5\% & -70.8\% & \underline{161.6\%} & -71.8\% & -67.3\% & -73.1\% & -73.0\% \\
\textbf{} & \textbf{} & \textbf{RW} & -12.3\% & -16.4\% & -34.3\% & -25.0\% & -12.3\% & -18.4\% & \underline{-81.1\%} & -35.6\% & \underline{-27.5\%} & 2.7\% & -38.6\% & -40.9\% & -11.9\% & \textbf{-89.9\%} & -60.3\% & -66.6\% & 132.1\% & -68.1\% & -62.0\% & -69.1\% & -82.1\% \\
\textbf{} & \textbf{} & \textbf{TS} & \textbf{-4.1\%} & \textbf{-6.6\%} & \textbf{-29.6\%} & \textbf{-8.4\%} & \textbf{-3.8\%} & \textbf{-4.3\%} & -69.7\% & \textbf{-17.1\%} & \textbf{-2.9\%} & 22.7\% & \textbf{-20.0\%} & \textbf{-22.3\%} & \textbf{17.9\%} & -85.0\% & \textbf{-48.5\%} & \textbf{-55.0\%} & 160.3\% & \textbf{-56.6\%} & \textbf{-51.1\%} & \textbf{-56.0\%} & -58.5\% \\
\rowfont{\color{ao(english)}}\textbf{} & \textbf{} & \textbf{*NS} & -42.8\% & -71.7\% & - & -77.1\% & -70.6\% & -74.9\% & - & -68.6\% & -69.4\% & \textbf{90.3\%} & -72.6\% & -72.6\% & -65.8\% & - & -92.2\% & -94.2\% & - & -94.0\% & -93.7\% & -93.5\% & - \\
\cline{2-24}
\textbf{} & \multirow[c]{6}{*}{\textbf{0.50}} & \textbf{FF} & -7.7\% & -10.5\% & -18.7\% & -15.0\% & -7.7\% & -10.8\% & \underline{-60.2\%} & -28.4\% & -29.5\% & 35.9\% & -32.2\% & -31.6\% & -26.9\% & -77.8\% & -46.5\% & -52.6\% & 170.6\% & -53.9\% & -48.7\% & -53.0\% & \underline{-67.9\%} \\
\textbf{} & \textbf{} & \textbf{FFB} & -6.7\% & -8.9\% & -32.7\% & -12.8\% & -6.3\% & -9.3\% & \textbf{-72.5\%} & -29.1\% & -29.1\% & 55.0\% & -32.2\% & -33.1\% & -25.8\% & \textbf{-82.5\%} & -47.3\% & -53.1\% & \textbf{266.1\%} & -53.7\% & -50.3\% & -52.2\% & -63.7\% \\
\textbf{} & \textbf{} & \textbf{PS} & \underline{-5.2\%} & \textbf{-4.9\%} & \underline{-16.9\%} & \underline{-9.1\%} & \underline{-3.2\%} & \textbf{-4.3\%} & -50.8\% & -30.5\% & -30.7\% & \textbf{72.4\%} & -33.5\% & -34.2\% & -28.6\% & -66.5\% & -49.9\% & -57.7\% & 205.7\% & -58.5\% & -52.7\% & -58.2\% & \textbf{-73.9\%} \\
\textbf{} & \textbf{} & \textbf{RJ} & \textbf{-5.1\%} & \underline{-5.9\%} & \textbf{-7.2\%} & -10.3\% & -4.4\% & -6.8\% & -17.4\% & \underline{-26.4\%} & -26.1\% & \underline{70.0\%} & \underline{-29.5\%} & \underline{-29.2\%} & \underline{-21.0\%} & -63.1\% & \textbf{-42.3\%} & \textbf{-50.2\%} & \underline{229.4\%} & \textbf{-49.8\%} & \textbf{-44.6\%} & \underline{-52.1\%} & -48.5\% \\
\textbf{} & \textbf{} & \textbf{RW} & -8.5\% & -16.6\% & -18.8\% & -17.0\% & -9.8\% & -13.6\% & -55.4\% & \textbf{-19.6\%} & \textbf{-15.7\%} & 63.1\% & \textbf{-21.7\%} & \textbf{-22.8\%} & \textbf{-8.2\%} & -61.2\% & -47.9\% & -55.4\% & 220.1\% & -56.8\% & -50.0\% & -55.2\% & -66.9\% \\
\textbf{} & \textbf{} & \textbf{TS} & -5.3\% & -6.1\% & -19.0\% & \textbf{-8.0\%} & \textbf{-2.2\%} & \underline{-4.3\%} & -55.1\% & -27.1\% & \underline{-25.6\%} & 59.4\% & -31.2\% & -30.5\% & -21.2\% & \underline{-80.7\%} & \underline{-42.9\%} & \underline{-50.6\%} & 226.2\% & \underline{-51.4\%} & \underline{-45.2\%} & \textbf{-50.9\%} & -57.3\% \\
\rowfont{\color{ao(english)}}\textbf{} & \textbf{} & \textbf{*NS} & -18.0\% & -42.0\% & -28.4\% & -47.0\% & -45.2\% & -53.1\% & - & -48.8\% & -47.5\% & 58.7\% & -52.2\% & -54.0\% & -41.3\% & - & -78.3\% & -84.3\% & - & -84.4\% & -81.0\% & -82.4\% & - \\
\cline{1-24} \cline{2-24}
\end{tabu}%
\begin{tabu}{RRZrrRRRRr|rrRRRRr|rrRRRRr|}
 &  &  & \multicolumn{21}{c|}{\textbf{INMO}} \\\hline
 &  &  & \multicolumn{7}{c|}{\textbf{ML-1M}} & \multicolumn{7}{c|}{\textbf{AB}} & \multicolumn{7}{c|}{\textbf{YD}} \\
 &  &  & \textbf{HR} & \textbf{NDCG} & \textbf{Cov} & \textbf{Re} & \textbf{Pr} & \textbf{AUC} & \textbf{Time} & \textbf{HR} & \textbf{NDCG} & \textbf{Cov} & \textbf{Re} & \textbf{Pr} & \textbf{AUC} & \textbf{Time} & \textbf{HR} & \textbf{NDCG} & \textbf{Cov} & \textbf{Re} & \textbf{Pr} & \textbf{AUC} & \textbf{Time} \\
\hline
\multirow[c]{25}{*}{\textbf{INMO}} & \textbf{-} & \textbf{-} 
& \textbf{0.663} & \textbf{0.185} & \textbf{0.338} & \textbf{0.090} & \textbf{0.158} & \textbf{0.101} & \textbf{1.2} 
& \textbf{0.154} & \textbf{0.038} & \textbf{0.252} & \textbf{0.072} & \textbf{0.010} & \textbf{0.017} & \textbf{7.3} 
& \textbf{0.171} & \textbf{0.033} & \textbf{0.110} & \textbf{0.055} & \textbf{0.012} & \textbf{0.015} & \textbf{45.5} \\
\cline{2-24}
\textbf{} & \multirow[c]{6}{*}{\textbf{0.20}} & \textbf{FF} & -5.8\% & -12.9\% & -71.5\% & -17.0\% & -11.7\% & -13.1\% & \textbf{-87.0\%} & -13.3\% & -19.3\% & -65.5\% & -17.8\% & -16.8\% & -20.6\% & -75.5\% & -20.5\% & -26.5\% & \underline{-66.7\%} & -26.8\% & -23.3\% & -27.3\% & -90.0\% \\
\textbf{} & \textbf{} & \textbf{FFB} & \underline{-3.6\%} & \underline{-6.6\%} & -54.2\% & \underline{-10.2\%} & \underline{-5.5\%} & -7.2\% & -85.2\% & -9.1\% & -11.3\% & \underline{-54.8\%} & -9.9\% & -10.8\% & -11.9\% & -74.5\% & -18.8\% & -24.5\% & -74.7\% & -24.5\% & -21.7\% & -24.3\% & \textbf{-94.2\%} \\
\textbf{} & \textbf{} & \textbf{PS} & \textbf{-1.4\%} & \textbf{-1.1\%} & \textbf{-36.1\%} & \textbf{-2.9\%} & \textbf{-0.7\%} & \textbf{-1.4\%} & -84.5\% & -6.9\% & -9.4\% & -56.6\% & -7.7\% & -9.7\% & -11.8\% & \textbf{-78.4\%} & \textbf{-15.5\%} & \underline{-20.2\%} & \textbf{-64.3\%} & \underline{-19.9\%} & \underline{-18.1\%} & \underline{-20.7\%} & -88.3\% \\
\textbf{} & \textbf{} & \textbf{RJ} & -3.7\% & -8.6\% & -60.7\% & -12.1\% & -7.5\% & -9.5\% & \underline{-86.8\%} & \underline{-6.2\%} & \underline{-6.3\%} & -55.1\% & \underline{-7.3\%} & \underline{-7.9\%} & \underline{-5.9\%} & -73.2\% & -22.2\% & -27.3\% & -71.8\% & -28.7\% & -25.2\% & -26.4\% & -92.6\% \\
\textbf{} & \textbf{} & \textbf{RW} & -3.6\% & -7.1\% & -57.7\% & -10.9\% & -6.0\% & -8.4\% & -84.7\% & \textbf{-3.8\%} & \textbf{-5.3\%} & -61.8\% & \textbf{-4.6\%} & \textbf{-6.2\%} & \textbf{-5.9\%} & \underline{-75.9\%} & -18.3\% & -24.4\% & -67.8\% & -24.7\% & -21.0\% & -25.7\% & -90.7\% \\
\textbf{} & \textbf{} & \textbf{TS} & -7.0\% & -7.6\% & \underline{-47.7\%} & -11.3\% & -6.9\% & \underline{-6.4\%} & -86.0\% & -14.2\% & -16.8\% & \textbf{-53.3\%} & -15.7\% & -15.2\% & -16.4\% & -75.1\% & \underline{-15.5\%} & \textbf{-19.5\%} & -69.2\% & \textbf{-19.9\%} & \textbf{-17.7\%} & \textbf{-17.5\%} & \underline{-93.2\%} \\
\rowfont{\color{ao(english)}}\textbf{} & \textbf{} & \textbf{*NS} & -3.8\% & -7.0\% & \textbf{18.4\%} & -10.9\% & -6.6\% & -10.5\% & - & -39.2\% & -43.1\% & \textbf{54.1\%} & -40.8\% & -43.4\% & -45.2\% & - & -24.3\% & -31.5\% & -75.8\% & -31.6\% & -28.8\% & -31.9\% & - \\
\cline{2-24}
\textbf{} & \multirow[c]{6}{*}{\textbf{0.50}} & \textbf{FF} & \underline{-0.4\%} & -3.0\% & -31.5\% & -4.0\% & -2.2\% & -2.2\% & -67.0\% & \underline{-1.8\%} & \underline{-0.6\%} & \underline{-24.9\%} & -2.4\% & \underline{-1.4\%} & 0.4\% & -69.7\% & -6.6\% & -10.0\% & -41.6\% & -9.9\% & -7.7\% & -11.0\% & \underline{-89.5\%} \\
\textbf{} & \textbf{} & \textbf{FFB} & \textbf{-0.4\%} & \textbf{-0.4\%} & -43.8\% & \textbf{-1.0\%} & \textbf{0.0\%} & \underline{-0.2\%} & \underline{-70.9\%} & -2.1\% & -1.0\% & \textbf{-24.0\%} & \underline{-1.5\%} & -2.0\% & 1.0\% & -73.3\% & \textbf{-4.4\%} & \underline{-6.0\%} & \textbf{-19.3\%} & \textbf{-6.0\%} & \underline{-5.2\%} & -6.2\% & -86.2\% \\
\textbf{} & \textbf{} & \textbf{PS} & -0.8\% & \underline{-0.4\%} & -30.8\% & \underline{-1.7\%} & \underline{-0.0\%} & \textbf{0.4\%} & -70.8\% & -5.6\% & -6.5\% & -29.1\% & -6.4\% & -6.2\% & -5.6\% & -75.1\% & -4.6\% & -6.3\% & -28.3\% & \underline{-6.4\%} & -5.2\% & \textbf{-5.4\%} & -87.5\% \\
\textbf{} & \textbf{} & \textbf{RJ} & -0.6\% & -2.0\% & \textbf{-27.0\%} & -3.1\% & -1.5\% & -2.2\% & -65.8\% & \textbf{-0.4\%} & \textbf{1.2\%} & -34.7\% & \textbf{-0.3\%} & \textbf{-0.8\%} & \underline{2.3\%} & \textbf{-80.1\%} & -5.2\% & -7.1\% & \underline{-24.0\%} & -7.4\% & -5.9\% & -7.5\% & -85.4\% \\
\textbf{} & \textbf{} & \textbf{RW} & -1.5\% & -3.2\% & \underline{-27.9\%} & -3.8\% & -3.1\% & -2.8\% & -63.6\% & -3.7\% & -0.8\% & -26.1\% & -4.3\% & -3.5\% & \textbf{2.4\%} & -64.7\% & \underline{-4.6\%} & \textbf{-5.8\%} & -27.6\% & -7.2\% & \textbf{-4.8\%} & \underline{-5.5\%} & -85.1\% \\
\textbf{} & \textbf{} & \textbf{TS} & -5.1\% & -5.6\% & -39.5\% & -8.1\% & -4.6\% & -4.6\% & \textbf{-73.7\%} & -10.4\% & -10.6\% & -25.6\% & -12.3\% & -10.6\% & -8.3\% & \underline{-78.7\%} & -5.8\% & -7.2\% & -31.8\% & -7.3\% & -6.9\% & -6.8\% & \textbf{-89.8\%} \\
\rowfont{\color{ao(english)}}\textbf{} & \textbf{} & \textbf{*NS} & -1.8\% & -4.0\% & \textbf{37.2\%} & -6.2\% & -4.3\% & -7.1\% & - & -19.1\% & -20.1\% & \textbf{40.1\%} & -20.6\% & -22.4\% & -19.7\% & - & -8.5\% & -11.9\% & -40.6\% & -11.2\% & -10.5\% & -11.0\% & - \\
\cline{1-24} \cline{2-24}
\end{tabu}%
\begin{tabu}{RRZrrRRRRr|rrRRRRr|rrRRRRr|}
 &  &  & \multicolumn{21}{c|}{\textbf{PinSAGE}} \\\hline
 &  &  & \multicolumn{7}{c|}{\textbf{ML-1M}} & \multicolumn{7}{c|}{\textbf{AB}} & \multicolumn{7}{c|}{\textbf{YD}} \\
 &  &  & \textbf{HR} & \textbf{NDCG} & \textbf{Cov} & \textbf{Re} & \textbf{Pr} & \textbf{AUC} & \textbf{Time} & \textbf{HR} & \textbf{NDCG} & \textbf{Cov} & \textbf{Re} & \textbf{Pr} & \textbf{AUC} & \textbf{Time} & \textbf{HR} & \textbf{NDCG} & \textbf{Cov} & \textbf{Re} & \textbf{Pr} & \textbf{AUC} & \textbf{Time} \\
\hline
\multirow[c]{25}{*}{\textbf{PinSAGE}} & \textbf{-} & \textbf{-} 
& \textbf{0.583} & \textbf{0.152} & \textbf{0.039} & \textbf{0.065} & \textbf{0.131} & \textbf{0.086} & \textbf{1.1} 
& \textbf{0.122} & \textbf{0.027} & \textbf{0.159} & \textbf{0.054} & \textbf{0.008} & \textbf{0.012} & \textbf{3.0} 
& \textbf{0.136} & \textbf{0.024} & \textbf{0.060} & \textbf{0.041} & \textbf{0.009} & \textbf{0.011} & \textbf{11.3} \\
\cline{2-24}
\textbf{} & \multirow[c]{6}{*}{\textbf{0.20}} & \textbf{FF} & -1.5\% & -15.3\% & -21.0\% & -7.6\% & -13.5\% & -20.5\% & \textbf{-87.7\%} & -26.6\% & -33.7\% & -36.5\% & -31.4\% & -28.6\% & -32.4\% & -85.3\% & -41.7\% & -50.7\% & -58.6\% & -50.7\% & -45.7\% & -53.6\% & \underline{-89.4\%} \\
\textbf{} & \textbf{} & \textbf{FFB} & \textbf{11.3\%} & \textbf{-0.0\%} & -9.9\% & \textbf{15.0\%} & 0.5\% & -10.7\% & -59.4\% & -27.3\% & -29.4\% & 6.0\% & -29.1\% & -28.8\% & \underline{-24.5\%} & -83.6\% & \underline{-30.2\%} & \underline{-34.5\%} & \underline{59.0\%} & \underline{-36.0\%} & \underline{-34.2\%} & \underline{-36.4\%} & -64.7\% \\
\textbf{} & \textbf{} & \textbf{PS} & -3.9\% & -3.4\% & \textbf{3.9\%} & -5.6\% & -2.9\% & \underline{-4.0\%} & -51.7\% & -25.5\% & -31.6\% & \textbf{29.7\%} & -32.0\% & -29.1\% & -26.9\% & \underline{-87.3\%} & -35.5\% & -42.3\% & 16.9\% & -41.3\% & -39.3\% & -42.9\% & -81.6\% \\
\textbf{} & \textbf{} & \textbf{RJ} & -4.9\% & -4.0\% & -16.0\% & -5.3\% & -1.6\% & -13.3\% & -74.0\% & \underline{-17.8\%} & \underline{-26.5\%} & \underline{14.8\%} & \underline{-22.9\%} & \underline{-23.5\%} & -24.9\% & -82.0\% & -41.1\% & -47.2\% & 19.6\% & -49.6\% & -45.3\% & -49.1\% & \textbf{-89.6\%} \\
\textbf{} & \textbf{} & \textbf{RW} & \underline{8.2\%} & \underline{-0.6\%} & -17.7\% & \underline{14.8\%} & \textbf{0.5\%} & -12.0\% & \underline{-79.1\%} & -45.5\% & -56.7\% & -40.9\% & -52.0\% & -50.2\% & -50.0\% & \textbf{-92.5\%} & -34.6\% & -41.9\% & \textbf{112.0\%} & -42.4\% & -37.3\% & -46.8\% & -69.8\% \\
\textbf{} & \textbf{} & \textbf{TS} & 6.7\% & -0.9\% & \underline{-8.3\%} & 3.9\% & \underline{0.5\%} & \textbf{-1.6\%} & -59.3\% & \textbf{-15.5\%} & \textbf{-21.0\%} & 9.5\% & \textbf{-20.3\%} & \textbf{-18.1\%} & \textbf{-17.9\%} & -86.1\% & \textbf{-20.5\%} & \textbf{-25.3\%} & 35.5\% & \textbf{-25.9\%} & \textbf{-23.7\%} & \textbf{-29.1\%} & -73.0\% \\
\rowfont{\color{ao(english)}}\textbf{} & \textbf{} & \textbf{*NS} & 1.8\% & -32.7\% & -52.5\% & -20.7\% & -36.3\% & -48.8\% & - & -46.0\% & -44.6\% & - & -47.8\% & -46.2\% & -38.5\% & - & -82.1\% & -86.5\% & \underline{78.5\%} & -85.5\% & -85.4\% & -86.8\% & - \\
\cline{2-24}
\textbf{} & \multirow[c]{6}{*}{\textbf{0.50}} & \textbf{FF} & -3.3\% & -3.2\% & -6.6\% & -3.1\% & 0.3\% & -5.7\% & \textbf{-55.0\%} & -20.7\% & -30.7\% & -22.6\% & -27.1\% & -23.0\% & -27.5\% & \textbf{-74.0\%} & -18.3\% & -21.8\% & 56.5\% & -22.7\% & -21.0\% & -22.6\% & \underline{-41.3\%} \\
\textbf{} & \textbf{} & \textbf{FFB} & \underline{9.2\%} & \underline{6.1\%} & -5.0\% & \underline{15.3\%} & \underline{5.9\%} & -0.8\% & 2.9\% & -18.4\% & -19.7\% & \underline{15.0\%} & -20.6\% & -18.3\% & -13.4\% & -55.6\% & \textbf{-14.8\%} & \textbf{-17.3\%} & 42.9\% & \underline{-19.4\%} & \textbf{-17.6\%} & \textbf{-19.0\%} & -6.1\% \\
\textbf{} & \textbf{} & \textbf{PS} & \textbf{13.2\%} & \textbf{6.4\%} & \textbf{2.2\%} & \textbf{21.2\%} & \textbf{6.1\%} & \textbf{2.0\%} & -14.3\% & -10.2\% & \underline{-13.8\%} & \textbf{25.1\%} & -14.0\% & -11.6\% & \underline{-10.3\%} & -4.0\% & -16.6\% & -19.3\% & 52.7\% & \textbf{-19.1\%} & -19.7\% & \underline{-20.2\%} & -16.3\% \\
\textbf{} & \textbf{} & \textbf{RJ} & 4.8\% & 2.0\% & \underline{-3.3\%} & 7.3\% & 3.5\% & -2.8\% & -26.1\% & \underline{-9.1\%} & -14.0\% & 12.6\% & \underline{-12.1\%} & \underline{-9.2\%} & -16.1\% & -49.4\% & \underline{-15.5\%} & \underline{-18.6\%} & \underline{71.4\%} & -19.9\% & \underline{-17.8\%} & -20.4\% & -18.5\% \\
\textbf{} & \textbf{} & \textbf{RW} & -0.1\% & 0.4\% & -3.9\% & 0.5\% & 0.7\% & -3.8\% & -6.4\% & -12.5\% & -15.9\% & 8.9\% & -15.1\% & -11.0\% & -10.4\% & -50.9\% & -16.7\% & -20.4\% & \textbf{96.5\%} & -20.7\% & -18.7\% & -24.1\% & -21.2\% \\
\textbf{} & \textbf{} & \textbf{TS} & 7.5\% & 2.8\% & -3.9\% & 5.9\% & 2.9\% & \underline{1.4\%} & \underline{-34.7\%} & \textbf{-8.3\%} & \textbf{-11.2\%} & 9.5\% & \textbf{-10.7\%} & \textbf{-8.0\%} & \textbf{-9.8\%} & \underline{-65.8\%} & -19.1\% & -23.8\% & -33.3\% & -24.1\% & -22.5\% & -25.8\% & \textbf{-70.1\%} \\
\rowfont{\color{ao(english)}}\textbf{} & \textbf{} & \textbf{*NS} & -21.0\% & -50.5\% & \textbf{2.2\%} & -57.2\% & -48.3\% & -52.7\% & - & -41.6\% & -46.0\% & \textbf{62.7\%} & -46.8\% & -44.1\% & -39.6\% & - & -70.1\% & -75.1\% & \underline{77.2\%} & -74.1\% & -75.4\% & -74.4\% & - \\
\cline{1-24} \cline{2-24}
\end{tabu}

%% file: tables/08_actionable_insights.tex
\begin{table*}[!htb]
\newcommand{\HY}{\hyphenpenalty=0\exhyphenpenalty=25\RaggedRight}
\centering
\caption{Summary of recommendations and insights for RQ1--RQ4}%
\label{tab:rq_summary}%
\vspace{-5pt}
\begin{tblr}{
    cells={valign=m,halign=l},
    row{1}={font=\bfseries,rowsep=0pt},
    column{1}={font=\bfseries, halign=c},
    rowsep=0pt,
    rightsep=0.5pt,
    leftsep=1pt,
    row{1-Z}={font=\footnotesize\color{ao(english)}},
    colspec={R >{\HY}m{.18\linewidth}| >{\HY}m{.23\linewidth}| >{\HY}m{.29\linewidth}| >{\HY}m{.27\linewidth}}
}
 & \textbf{RQ1: Samplers} & \textbf{RQ2: Sampling ratio} & \textbf{RQ3: Training time} & \textbf{RQ4: Recommenders} \\ \hline
\textbf{Combined}
&

    TS/PS perform best in most settings.

    For ratios ${\leq}0.10$ use TS.

    Samplers matching the degree distribution perform better.
&
    $r{=}0.5$ maintains recommendation performance with great time savings.


    $r{\leq}0.10$ does not maintain performance on datasets with less than 2.5mil ratings unless popularity biased.
&
    ~80\% time reduction at $r=0.5$ for GNN-based recommender without neighborhood sampling.

    Attention requires longer time for apt performance.

    Subsampling ${\geq}0.20$ for hyperparameter tuning is possible.

&
    Learned features are superior in most settings.

    When side information or KG features are key to performance, ensure enough data is retained to cover relevant entities.

    Methods learning collaborative bias are less dependent on the samplers.
\end{tblr}
\vspace{-5pt}
\end{table*}

%% file: sections/99_conclusion.tex
We investigate the practical implications of employing subsampled training data with different graph-based sampling methodologies when training inductive \glspl{rs}.
Inductive techniques are able to provide predictions for out-of-samples data and past evaluations have exploited this ability to reduce training time.
Our evaluation shows that the PinSAGE and Temporal sampling approach produces the most reliable samples.
Yet, for all \glspl{rs}, at least $50\%$ of the training graph is required to maintain prediction accuracy except for popularity biased datasets where $\leq10\%$ is sufficient.
Nonetheless, the most robust method, INMO, showcases an important reduction in training time, with a decrease of over $80\%$, already with $50\%$ of the graph. 
Future research could design sampling methods for and increase the robustness of inductive recommendation methods.

%% file: appendix/00_index.tex
\input{figures/XA_01_skews}%
\input{tables/XA_02_hp}%
\input{tables/XA_03_toppop}%

~
\newpage

\section{Extended result}\label{app:results}
\input{appendix/01_results}

\section{Hyperparameter tuning}\label{app:hp}
\input{appendix/02_hyperparameter}

\section{TopPop comparison}\label{app:tp}
\input{appendix/03_toppop}

\section{User ratio and rating skew}\label{app:urrs}
\input{appendix/04_ratio_skew}

\input{tables/XA_01_results}

%% file: figures/XA_01_skews.tex
\begin{figure}[tb]
        \centering
        \includegraphics[height=.33cm]{figures/graph_statistic/graph_statistic_legend.pdf}\\
    
        \settoheight{\imageheight}{\includegraphics[width=0.5\linewidth, trim=0 0 0 0]{figures/graph_statistic/YD_IV_graph_statistic.pdf}}

        \includegraphics[height=\imageheight, trim=30 0 0 0, clip]{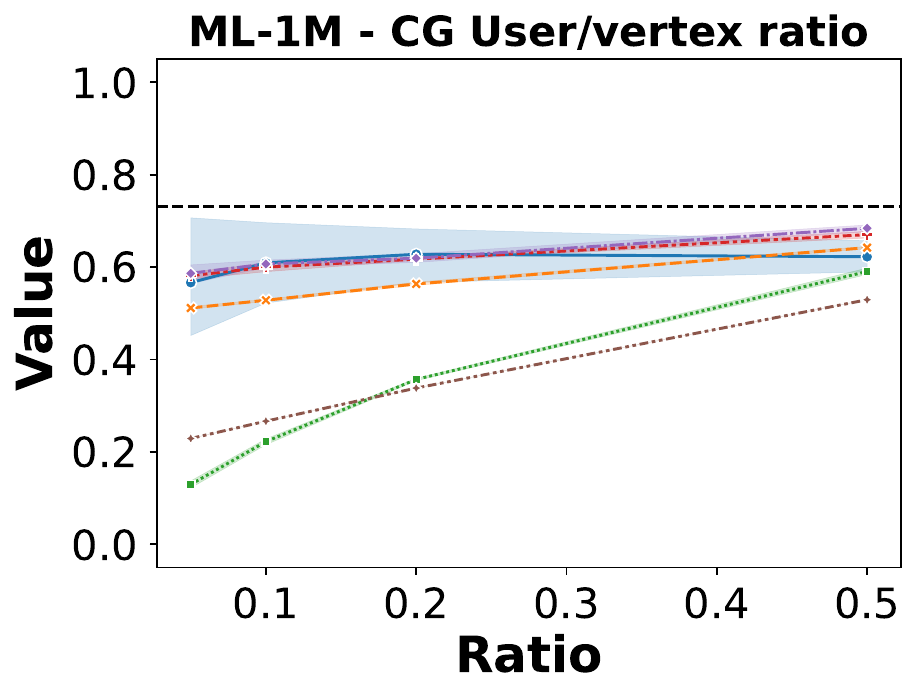}%
        \includegraphics[height=\imageheight, trim=30 0 0 0, clip]{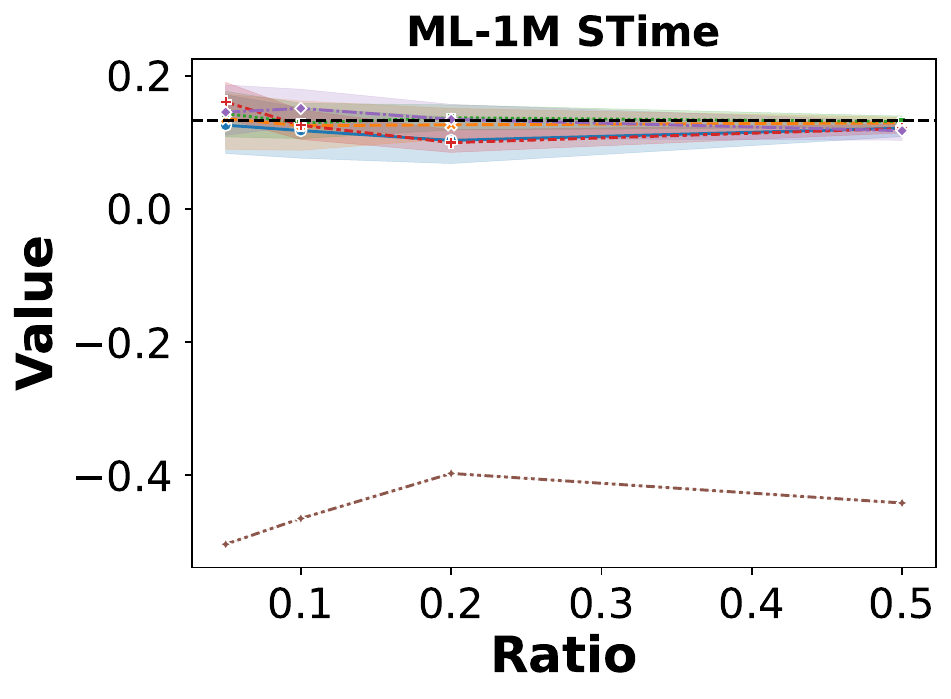}%

        \includegraphics[height=\imageheight, trim=30 0 0 0, clip]{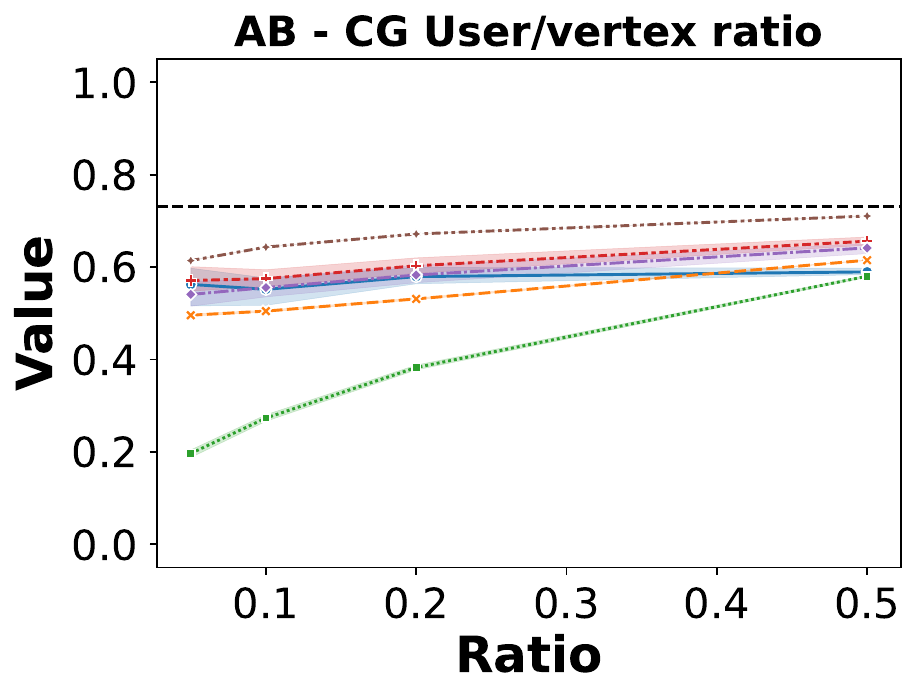}%
        \includegraphics[height=\imageheight, trim=30 0 0 0, clip]{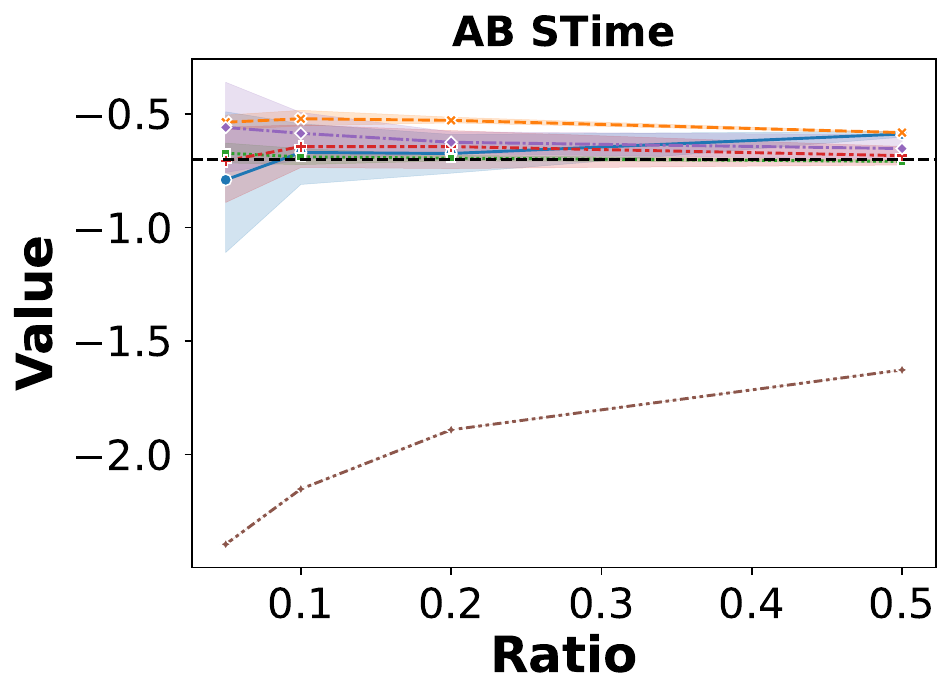}%
        
        \includegraphics[height=\imageheight, trim=30 0 0 0, clip]{figures/graph_statistic/YD_UV_graph_statistic.pdf}%
        \includegraphics[height=\imageheight, trim=30 0 0 0, clip]{figures/graph_statistic/YD_STime_graph_statistic.pdf}%
        \vspace{-10pt}
        \caption{Illustration of user ratio and rating skew in the YD.}
        \label{figapp:ratioskew}
\end{figure}%

%% file: tables/XA_02_hp.tex
\begin{table}[tb]
    \centering
    \resizebox{1\linewidth}{!}{%
    \begin{tabular}{lll|rr|rr|rr}
         &  &  & \multicolumn{2}{c|}{\bfseries MovieLens} & \multicolumn{2}{c|}{\bfseries Amazon Book} & \multicolumn{2}{c}{\bfseries Yelp} \\
         &  &  & \bfseries HR & \bfseries NDCG & \bfseries HR & \bfseries NDCG & \bfseries HR & \bfseries NDCG \\
        \hline
        \multirow[c]{4}{*}{\bfseries GInRec} & \multirow[c]{2}{*}{\bfseries 0.2} & \bfseries Base & 0.683 & 0.185 & 0.081 & 0.021 & 0.047 & 0.006 \\
        \bfseries  & \bfseries  & \bfseries HP & \bfseries 0.757 & \bfseries 0.205 & \bfseries 0.106 & \bfseries 0.026 & \bfseries 0.081 & \bfseries 0.013 \\
        \cline{2-9}
        \bfseries  & \multirow[c]{2}{*}{\bfseries 0.5} & \bfseries Base & 0.696 & 0.193 & 0.092 & 0.020 & 0.061 & 0.009 \\
        \bfseries  & \bfseries  & \bfseries HP & \bfseries 0.723 & \bfseries 0.195 & \bfseries 0.117 & \bfseries 0.028 & \bfseries 0.101 & \bfseries 0.017 \\
        \cline{1-9} \cline{2-9}
        \multirow[c]{4}{*}{\bfseries INMO} & \multirow[c]{2}{*}{\bfseries 0.2} & \bfseries Base & \bfseries 0.653 & 0.182 & \bfseries 0.144 & \bfseries 0.035 & 0.144 & 0.027 \\
        \bfseries  & \bfseries  & \bfseries HP & 0.651 & \bfseries 0.184 & 0.141 & 0.035 & \bfseries 0.147 & \bfseries 0.027 \\
        \cline{2-9}
        \bfseries  & \multirow[c]{2}{*}{\bfseries 0.5} & \bfseries Base & \bfseries 0.658 & 0.184 & \bfseries 0.146 & \bfseries 0.036 & \bfseries 0.163 & \bfseries 0.031 \\
        \bfseries  & \bfseries  & \bfseries HP & 0.653 & \bfseries 0.185 & 0.141 & 0.036 & 0.161 & 0.031 \\
        \cline{1-9} \cline{2-9}
        \multirow[c]{4}{*}{\bfseries PinSAGE} & \multirow[c]{2}{*}{\bfseries 0.2} & \bfseries Base & 0.561 & 0.147 & 0.091 & 0.019 & 0.088 & 0.014 \\
        \bfseries  & \bfseries  & \bfseries HP & \bfseries 0.627 & \bfseries 0.155 & \bfseries 0.113 & \bfseries 0.026 & \bfseries 0.097 & \bfseries 0.017 \\
        \cline{2-9}
        \bfseries  & \multirow[c]{2}{*}{\bfseries 0.5} & \bfseries Base & \bfseries 0.661 & \bfseries 0.161 & \bfseries 0.110 & 0.024 & 0.113 & 0.020 \\
        \bfseries  & \bfseries  & \bfseries HP & 0.635 & 0.160 & 0.102 & \bfseries 0.025 & \bfseries 0.123 & \bfseries 0.022 \\
        \cline{1-9} \cline{2-9}
    \end{tabular}%
    }
    \caption{Hyperparameter tuning performed with PS. HP is hyperparameter tuned and Base are the results of \autoref{tabapp:results}.}
    \label{tabapp:hp}
\end{table}%

%% file: tables/XA_03_toppop.tex
\begin{table}[tb]
    \centering
    \resizebox{.8\linewidth}{!}{%
    \begin{tabular}{r|rrZZZZ|rrZZZZ|rrZZZZ}
         & \multicolumn{6}{c|}{\textbf{MovieLens}} & \multicolumn{6}{c|}{\textbf{Amazon Book}} & \multicolumn{6}{c}{\textbf{Yelp}} \\ 
         & \multicolumn{1}{c}{\textbf{Hr}} & \multicolumn{1}{c}{\textbf{NDCG}} & \multicolumn{1}{Z}{\textbf{Recall}} & \multicolumn{1}{Z}{\textbf{Precision}} & \multicolumn{1}{Z}{\textbf{Cov}} & \multicolumn{1}{Z|}{\textbf{AUC}} & \multicolumn{1}{c}{\textbf{Hr}} & \multicolumn{1}{c}{\textbf{NDCG}} & \multicolumn{1}{Z}{\textbf{Recall}} & \multicolumn{1}{Z}{\textbf{Precision}} & \multicolumn{1}{Z}{\textbf{Cov}} & \multicolumn{1}{Z|}{\textbf{AUC}} & \multicolumn{1}{c}{\textbf{Hr}} & \multicolumn{1}{c}{\textbf{NDCG}} & \multicolumn{1}{Z}{\textbf{Recall}} & \multicolumn{1}{Z}{\textbf{Precision}} & \multicolumn{1}{Z}{\textbf{Cov}} & \multicolumn{1}{Z}{\textbf{AUC}} \\ \hline
        \textbf{TopPop} 
            & {0.601} & {0.134} & {0.064} & {0.117} & {0.034} & {0.076} 
            & {0.028} & {0.006} & {0.010} & {0.002} & {0.002} & {0.003} 
            & {0.033} & {0.005} & {0.009} & {0.002} & {0.000} & {0.002} \\
        \textbf{GInRec} 
            & \textbf{0.668} & \textbf{0.173} & 0.119 & {0.082} & {0.151} & {0.093} 
            & {0.086} & \underline{0.023} & 0.117 & {0.036} & {0.005} & {0.012} 
            & 0.014 & 0.002 & 0.035 & 0.003 & 0.001 & 0.001 \\
        \textbf{INMO} 
            & {0.582} & \underline{0.158} & {0.061} & {0.069} & {0.137} & {0.086}  
            & \textbf{0.135} & \textbf{0.032} & 0.021 & {0.062} & {0.008} & {0.014}  
            & \textbf{0.119} & \textbf{0.021} & 0.011 & {0.035} & {0.008} & {0.009} \\
        \textbf{PinSAGE} 
            & \underline{0.661} & 0.148 & {0.031} & {0.070} & {0.128} & {0.077}  
            & \underline{0.097} & {0.020} & 0.139 & {0.040} & {0.006} & {0.009}  
            & \underline{0.080} & \underline{0.013} & {0.096} & {0.021} & {0.005} & {0.005} \\
    \end{tabular}%
    }
    \caption{Performance of methods using TS at $r=0.05$ compared to TopPop}
    \label{tabapp:tp}
\end{table}%

%% file: appendix/01_results.tex
We compute a lot of metrics HR, NDCG, Coverage, Recall, Precision and AUC, not included in the full paper.
We show the remaining results in \autoref{tabapp:results}. 
Interestingly TS is often less good at Coverage, but is otherwise well performing a cross all metrics, with particularly PinSAGE preferring TS for $r\leq0.20$. 

We also note that in most cases NDCG and HR show the same relative ordering as in other metrics, with only Coverage showing different ratios. 
However, Coverage does not measure the ranking ability.
GInRecs performance decrease at 50\% for YD is likely due to very high coverage, meaning more random recommendations in this case.

%% file: appendix/02_hyperparameter.tex
In \autoref{tabapp:hp} we show that the methods are capable of hyperparameter tune with ratio's $0.2$ and $0.5$, illustrating that subsampling is feasible for both training and tuning. 
GInRec for example gains twice the performance on Yelp, increasing in percent points is 33-38\% for $0.2$ and 17-19\% for $0.5$, relative to the GInRec's performance on the full graph.
This indicates that GInRec is more sensitive to the hyperparameters than other methods. 
In contrast, INMO and PinSAGE's performance do not change significantly; however, some improvements can be seen for PinSAGE at $r=0.2$.



%% file: appendix/03_toppop.tex
In \autoref{tabapp:tp} we show that all methods are capable of outperforming TopPop except GInRec on Yelp and HR of INMO on the ML dataset. 
However, based on the results of \autoref{app:hp}, it is likely that the performance of GInRec could be improved beyond that of TopPop with hyperparameter tuning.
With high popularity, it may not make sense to train as TopPop selects a lot of relevant items for the ML dataset, but for AB and YD we see between $260-433\%$ increase in performance for INMO.
Thus, for more difficult datasets it makes sense to train a method even with low ratios.

%% file: appendix/04_ratio_skew.tex
In \autoref{figapp:ratioskew}, we show the user ratio and STime for all datasets. 
We see that TS has the most different STime for all dataset and that PS's user ratio is often very different from the other datasets.

%% file: tables/XA_01_results.tex
\begin{table*}[tb]
    \centering
    \resizebox{1\linewidth}{!}{%
    \input{tables/results/sampling_rel_mrs}
    }
    \caption{All results of the methods.}
    \label{tabapp:results}
\end{table*}

%% file: tables/results/sampling_rel_mrs.tex
\begin{tabular}{lll|rrrrrrr|rrrrrrr|rrrrrrr}
 &  &  & \multicolumn{7}{c}{\textbf{ML-1M}} & \multicolumn{7}{c}{\textbf{AB}} & \multicolumn{7}{c}{\textbf{YD}} \\
 &  &  & \textbf{HR} & \textbf{NDCG} & \textbf{Cov} & \textbf{Re} & \textbf{Pr} & \textbf{AUC} & \textbf{Time} & \textbf{HR} & \textbf{NDCG} & \textbf{Cov} & \textbf{Re} & \textbf{Pr} & \textbf{AUC} & \textbf{Time} & \textbf{HR} & \textbf{NDCG} & \textbf{Cov} & \textbf{Re} & \textbf{Pr} & \textbf{AUC} & \textbf{Time} \\
\hline
\multirow[c]{27}{*}{\textbf{GInRec}} & \textbf{-} & \textbf{-} & \textbf{0.734} & \textbf{0.202} & \textbf{0.355} & \textbf{0.103} & \textbf{0.168} & \textbf{0.111} & \textbf{0.898} & \textbf{0.132} & \textbf{0.029} & \textbf{0.390} & \textbf{0.059} & \textbf{0.008} & \textbf{0.013} & \textbf{2.732} & \textbf{0.122} & \textbf{0.021} & \textbf{0.135} & \textbf{0.036} & \textbf{0.008} & \textbf{0.009} & \textbf{7.915} \\
\cline{2-24}
\textbf{} & \multirow[c]{6}{*}{\textbf{0.05}} & \textbf{FF} & -16.8\% & -25.1\% & \textbf{-51.9\%} & -35.0\% & -19.8\% & -30.4\% & -82.3\% & \underline{-45.6\%} & -48.2\% & \underline{-31.0\%} & \underline{-50.3\%} & \underline{-49.9\%} & -46.2\% & -86.4\% & \underline{-83.1\%} & \textbf{-85.8\%} & \underline{-19.8\%} & \textbf{-85.9\%} & \underline{-85.1\%} & \underline{-87.3\%} & -85.4\% \\
\textbf{} & \textbf{} & \textbf{FFB} & -17.8\% & -25.1\% & -70.5\% & -37.4\% & -20.3\% & -29.6\% & \textbf{-90.8\%} & -67.5\% & -67.5\% & -63.9\% & -69.0\% & -72.3\% & -64.1\% & -92.9\% & -89.0\% & -90.0\% & -77.0\% & -90.6\% & -91.1\% & -89.2\% & -95.2\% \\
\textbf{} & \textbf{} & \textbf{PS} & \textbf{-5.1\%} & \textbf{-8.4\%} & -64.5\% & \textbf{-13.4\%} & \textbf{-6.5\%} & \textbf{-10.7\%} & -81.1\% & -47.8\% & -55.2\% & \textbf{-26.7\%} & -51.8\% & -54.1\% & -57.3\% & -91.9\% & -88.2\% & -89.6\% & -68.4\% & -90.2\% & -90.4\% & -89.4\% & \underline{-95.3\%} \\
\textbf{} & \textbf{} & \textbf{RJ} & -15.6\% & -25.9\% & \underline{-63.5\%} & -33.5\% & -20.3\% & -31.3\% & -89.1\% & -63.2\% & -66.2\% & -61.6\% & -65.2\% & -68.8\% & -68.2\% & \textbf{-94.0\%} & -88.4\% & -90.3\% & -31.1\% & -90.3\% & -90.3\% & -90.5\% & -94.0\% \\
\textbf{} & \textbf{} & \textbf{RW} & -11.3\% & -20.9\% & -68.5\% & -29.7\% & -17.1\% & -28.5\% & -87.5\% & -64.2\% & \underline{-47.6\%} & -56.4\% & -64.1\% & -70.1\% & \underline{-22.6\%} & -91.7\% & \textbf{-82.3\%} & \underline{-85.9\%} & \textbf{42.7\%} & \underline{-86.8\%} & \textbf{-84.6\%} & \textbf{-86.9\%} & -83.4\% \\
\textbf{} & \textbf{} & \textbf{TS} & \underline{-8.9\%} & \underline{-14.4\%} & -66.5\% & \underline{-19.9\%} & \underline{-10.0\%} & \underline{-15.8\%} & \underline{-89.7\%} & \textbf{-34.9\%} & \textbf{-22.0\%} & -69.9\% & \textbf{-38.6\%} & \textbf{-42.0\%} & \textbf{-3.4\%} & \underline{-93.5\%} & -88.4\% & -90.1\% & -74.0\% & -91.0\% & -90.6\% & -89.9\% & \textbf{-95.6\%} \\
\cline{2-24}
\textbf{} & \multirow[c]{6}{*}{\textbf{0.10}} & \textbf{FF} & -10.8\% & -15.9\% & \underline{-41.8\%} & -23.3\% & -12.7\% & -20.9\% & -78.9\% & -44.6\% & -44.1\% & -18.0\% & -48.4\% & -49.5\% & -37.5\% & -82.4\% & -76.3\% & -79.2\% & 46.3\% & -79.9\% & -78.1\% & -80.9\% & -83.1\% \\
\textbf{} & \textbf{} & \textbf{FFB} & -9.5\% & -15.3\% & -44.0\% & -20.0\% & -11.3\% & -19.2\% & -82.6\% & -42.4\% & -36.3\% & 4.1\% & -45.2\% & -48.1\% & -23.1\% & -82.0\% & \underline{-71.5\%} & \underline{-76.7\%} & \underline{69.1\%} & \underline{-77.2\%} & \underline{-73.8\%} & \underline{-79.2\%} & -80.1\% \\
\textbf{} & \textbf{} & \textbf{PS} & \textbf{-1.3\%} & \textbf{-1.8\%} & -46.8\% & \textbf{-5.0\%} & \textbf{-3.5\%} & \textbf{-5.2\%} & -76.8\% & -49.0\% & -47.0\% & -1.0\% & -50.7\% & -55.3\% & -39.7\% & \textbf{-87.0\%} & \textbf{-68.0\%} & \textbf{-74.7\%} & 3.2\% & \textbf{-75.8\%} & \textbf{-71.7\%} & \textbf{-76.1\%} & -77.9\% \\
\textbf{} & \textbf{} & \textbf{RJ} & -9.7\% & -16.1\% & \textbf{-31.8\%} & -22.0\% & -12.1\% & -20.7\% & -72.0\% & \underline{-32.6\%} & \underline{-21.4\%} & \textbf{7.5\%} & \underline{-34.1\%} & \underline{-39.6\%} & -6.5\% & -79.2\% & -77.9\% & -81.7\% & \textbf{129.2\%} & -81.9\% & -79.6\% & -83.1\% & -81.7\% \\
\textbf{} & \textbf{} & \textbf{RW} & -11.7\% & -20.8\% & -63.9\% & -27.7\% & -15.4\% & -25.3\% & \textbf{-89.8\%} & -45.9\% & -30.3\% & -3.7\% & -47.2\% & -51.8\% & \underline{-3.4\%} & -83.5\% & -76.8\% & -81.1\% & 36.8\% & -81.9\% & -79.1\% & -83.7\% & \underline{-91.2\%} \\
\textbf{} & \textbf{} & \textbf{TS} & \underline{-7.9\%} & \underline{-11.4\%} & -60.1\% & \underline{-15.4\%} & \underline{-7.2\%} & \underline{-11.5\%} & \underline{-89.0\%} & \textbf{-18.5\%} & \textbf{-7.8\%} & \underline{4.2\%} & \textbf{-21.6\%} & \textbf{-23.4\%} & \textbf{8.3\%} & \underline{-85.6\%} & -86.7\% & -88.4\% & -64.8\% & -89.2\% & -89.1\% & -88.5\% & \textbf{-95.5\%} \\
\cline{2-24}
\textbf{} & \multirow[c]{7}{*}{\textbf{0.20}} & \textbf{FF} & \underline{-5.1\%} & -10.3\% & -33.6\% & \underline{-12.1\%} & -7.9\% & -12.8\% & -75.5\% & -34.2\% & -33.9\% & -5.3\% & -39.2\% & -38.3\% & -28.0\% & -82.6\% & -62.5\% & -68.0\% & 94.8\% & -68.7\% & -64.7\% & -70.3\% & \textbf{-84.3\%} \\
\textbf{} & \textbf{} & \textbf{FFB} & -9.5\% & -12.6\% & -32.6\% & -18.5\% & -9.3\% & -14.1\% & -78.0\% & \underline{-31.0\%} & -30.1\% & \underline{44.9\%} & \underline{-34.5\%} & \underline{-35.7\%} & -24.5\% & -79.5\% & \underline{-58.2\%} & \underline{-64.0\%} & \underline{174.5\%} & \underline{-65.6\%} & \underline{-60.4\%} & \underline{-65.4\%} & -61.6\% \\
\textbf{} & \textbf{} & \textbf{NS} & -42.8\% & -71.7\% & \underline{-30.1\%} & -77.1\% & -70.6\% & -74.9\% & \textbf{-86.3\%} & -68.6\% & -69.4\% & \textbf{90.3\%} & -72.6\% & -72.6\% & -65.8\% & -86.1\% & -92.2\% & -94.2\% & \textbf{279.9\%} & -94.0\% & -93.7\% & -93.5\% & -78.1\% \\
\textbf{} & \textbf{} & \textbf{PS} & -6.9\% & \underline{-8.7\%} & -31.5\% & -14.5\% & \underline{-6.1\%} & \underline{-9.8\%} & -68.4\% & -38.8\% & -29.3\% & 29.5\% & -42.1\% & -45.2\% & \underline{-10.7\%} & \underline{-89.1\%} & -61.5\% & -68.9\% & 61.1\% & -69.1\% & -65.0\% & -70.2\% & \underline{-82.5\%} \\
\textbf{} & \textbf{} & \textbf{RJ} & -10.0\% & -15.7\% & -45.0\% & -18.7\% & -11.6\% & -17.5\% & \underline{-83.3\%} & -35.8\% & -37.6\% & 20.5\% & -39.7\% & -40.7\% & -34.1\% & -81.6\% & -64.5\% & -70.8\% & 161.6\% & -71.8\% & -67.3\% & -73.1\% & -73.0\% \\
\textbf{} & \textbf{} & \textbf{RW} & -12.3\% & -16.4\% & -34.3\% & -25.0\% & -12.3\% & -18.4\% & -81.1\% & -35.6\% & \underline{-27.5\%} & 2.7\% & -38.6\% & -40.9\% & -11.9\% & \textbf{-89.9\%} & -60.3\% & -66.6\% & 132.1\% & -68.1\% & -62.0\% & -69.1\% & -82.1\% \\
\textbf{} & \textbf{} & \textbf{TS} & \textbf{-4.1\%} & \textbf{-6.6\%} & \textbf{-29.6\%} & \textbf{-8.4\%} & \textbf{-3.8\%} & \textbf{-4.3\%} & -69.7\% & \textbf{-17.1\%} & \textbf{-2.9\%} & 22.7\% & \textbf{-20.0\%} & \textbf{-22.3\%} & \textbf{17.9\%} & -85.0\% & \textbf{-48.5\%} & \textbf{-55.0\%} & 160.3\% & \textbf{-56.6\%} & \textbf{-51.1\%} & \textbf{-56.0\%} & -58.5\% \\
\cline{2-24}
\textbf{} & \multirow[c]{7}{*}{\textbf{0.50}} & \textbf{FF} & -7.7\% & -10.5\% & -18.7\% & -15.0\% & -7.7\% & -10.8\% & -60.2\% & -28.4\% & -29.5\% & 35.9\% & -32.2\% & -31.6\% & -26.9\% & -77.8\% & -46.5\% & -52.6\% & 170.6\% & -53.9\% & -48.7\% & -53.0\% & \underline{-67.9\%} \\
\textbf{} & \textbf{} & \textbf{FFB} & -6.7\% & -8.9\% & -32.7\% & -12.8\% & -6.3\% & -9.3\% & \textbf{-72.5\%} & -29.1\% & -29.1\% & 55.0\% & -32.2\% & -33.1\% & -25.8\% & \underline{-82.5\%} & -47.3\% & -53.1\% & \underline{266.1\%} & -53.7\% & -50.3\% & -52.2\% & -63.7\% \\
\textbf{} & \textbf{} & \textbf{NS} & -18.0\% & -42.0\% & -28.4\% & -47.0\% & -45.2\% & -53.1\% & \underline{-63.5\%} & -48.8\% & -47.5\% & 58.7\% & -52.2\% & -54.0\% & -41.3\% & \textbf{-89.6\%} & -78.3\% & -84.3\% & \textbf{290.5\%} & -84.4\% & -81.0\% & -82.4\% & -31.2\% \\
\textbf{} & \textbf{} & \textbf{PS} & \underline{-5.2\%} & \textbf{-4.9\%} & \underline{-16.9\%} & \underline{-9.1\%} & \underline{-3.2\%} & \textbf{-4.3\%} & -50.8\% & -30.5\% & -30.7\% & \textbf{72.4\%} & -33.5\% & -34.2\% & -28.6\% & -66.5\% & -49.9\% & -57.7\% & 205.7\% & -58.5\% & -52.7\% & -58.2\% & \textbf{-73.9\%} \\
\textbf{} & \textbf{} & \textbf{RJ} & \textbf{-5.1\%} & \underline{-5.9\%} & \textbf{-7.2\%} & -10.3\% & -4.4\% & -6.8\% & -17.4\% & \underline{-26.4\%} & -26.1\% & \underline{70.0\%} & \underline{-29.5\%} & \underline{-29.2\%} & \underline{-21.0\%} & -63.1\% & \textbf{-42.3\%} & \textbf{-50.2\%} & 229.4\% & \textbf{-49.8\%} & \textbf{-44.6\%} & \underline{-52.1\%} & -48.5\% \\
\textbf{} & \textbf{} & \textbf{RW} & -8.5\% & -16.6\% & -18.8\% & -17.0\% & -9.8\% & -13.6\% & -55.4\% & \textbf{-19.6\%} & \textbf{-15.7\%} & 63.1\% & \textbf{-21.7\%} & \textbf{-22.8\%} & \textbf{-8.2\%} & -61.2\% & -47.9\% & -55.4\% & 220.1\% & -56.8\% & -50.0\% & -55.2\% & -66.9\% \\
\textbf{} & \textbf{} & \textbf{TS} & -5.3\% & -6.1\% & -19.0\% & \textbf{-8.0\%} & \textbf{-2.2\%} & \underline{-4.3\%} & -55.1\% & -27.1\% & \underline{-25.6\%} & 59.4\% & -31.2\% & -30.5\% & -21.2\% & -80.7\% & \underline{-42.9\%} & \underline{-50.6\%} & 226.2\% & \underline{-51.4\%} & \underline{-45.2\%} & \textbf{-50.9\%} & -57.3\% \\
\cline{1-24} \cline{2-24}
\multirow[c]{27}{*}{\textbf{INMO}} & \textbf{-} & \textbf{-} & \textbf{0.663} & \textbf{0.185} & \textbf{0.338} & \textbf{0.090} & \textbf{0.158} & \textbf{0.101} & \textbf{1.248} & \textbf{0.154} & \textbf{0.038} & \textbf{0.252} & \textbf{0.072} & \textbf{0.010} & \textbf{0.017} & \textbf{7.291} & \textbf{0.171} & \textbf{0.033} & \textbf{0.110} & \textbf{0.055} & \textbf{0.012} & \textbf{0.015} & \textbf{45.522} \\
\cline{2-24}
\textbf{} & \multirow[c]{6}{*}{\textbf{0.05}} & \textbf{FF} & -17.0\% & -22.6\% & -88.8\% & -30.6\% & -21.9\% & -22.5\% & -96.5\% & -35.3\% & -49.4\% & -92.8\% & -37.0\% & -41.0\% & -59.2\% & -81.6\% & -49.8\% & -57.3\% & -92.8\% & -58.2\% & -52.0\% & -57.4\% & -93.0\% \\
\textbf{} & \textbf{} & \textbf{FFB} & -16.9\% & -22.3\% & -88.7\% & -30.6\% & -21.7\% & -22.3\% & -96.7\% & -22.7\% & -33.8\% & -91.3\% & -27.7\% & -28.4\% & -38.8\% & -80.9\% & -40.9\% & -47.4\% & -92.7\% & -48.6\% & -44.2\% & -46.8\% & \underline{-96.0\%} \\
\textbf{} & \textbf{} & \textbf{PS} & \textbf{-11.9\%} & \underline{-14.8\%} & \underline{-84.1\%} & \textbf{-22.5\%} & \underline{-13.7\%} & \textbf{-14.2\%} & -93.0\% & \underline{-15.7\%} & \underline{-26.2\%} & \textbf{-89.1\%} & \underline{-17.7\%} & \underline{-20.7\%} & \underline{-34.9\%} & -83.4\% & -34.2\% & -40.8\% & \underline{-89.5\%} & -41.2\% & -37.4\% & -39.5\% & -89.2\% \\
\textbf{} & \textbf{} & \textbf{RJ} & -16.9\% & -22.2\% & -88.7\% & -30.7\% & -21.7\% & -22.1\% & \underline{-97.0\%} & -23.4\% & -36.6\% & \underline{-90.8\%} & -27.9\% & -29.0\% & -44.4\% & \underline{-84.0\%} & \underline{-33.5\%} & \underline{-39.9\%} & \textbf{-88.0\%} & \underline{-40.8\%} & \underline{-36.8\%} & \underline{-39.0\%} & -91.2\% \\
\textbf{} & \textbf{} & \textbf{RW} & -16.8\% & -22.4\% & -88.6\% & -30.6\% & -21.7\% & -22.3\% & \textbf{-97.4\%} & -23.9\% & -37.4\% & -92.0\% & -26.0\% & -29.4\% & -49.4\% & -82.3\% & -52.5\% & -59.4\% & -94.6\% & -60.0\% & -55.0\% & -60.0\% & \textbf{-96.1\%} \\
\textbf{} & \textbf{} & \textbf{TS} & \underline{-12.1\%} & \textbf{-14.6\%} & \textbf{-81.8\%} & \underline{-23.1\%} & \textbf{-13.4\%} & \underline{-14.8\%} & -92.7\% & \textbf{-12.6\%} & \textbf{-15.9\%} & -91.6\% & \textbf{-13.9\%} & \textbf{-16.8\%} & \textbf{-18.3\%} & \textbf{-86.9\%} & \textbf{-30.1\%} & \textbf{-37.2\%} & -90.3\% & \textbf{-36.9\%} & \textbf{-33.2\%} & \textbf{-36.3\%} & -95.1\% \\
\cline{2-24}
\textbf{} & \multirow[c]{6}{*}{\textbf{0.10}} & \textbf{FF} & -13.6\% & -19.5\% & -80.7\% & -28.4\% & -18.2\% & -18.9\% & \textbf{-91.8\%} & -29.6\% & -38.1\% & -87.9\% & -32.5\% & -35.3\% & -42.8\% & -86.4\% & -41.8\% & -49.9\% & -82.7\% & -50.4\% & -44.9\% & -51.2\% & -90.9\% \\
\textbf{} & \textbf{} & \textbf{FFB} & -6.9\% & -11.4\% & -75.0\% & -16.2\% & -9.8\% & -13.0\% & -91.6\% & -17.0\% & -23.0\% & \underline{-79.4\%} & -20.4\% & -20.7\% & -24.2\% & -79.3\% & -52.4\% & -59.4\% & -93.7\% & -60.3\% & -56.8\% & -58.6\% & \textbf{-99.2\%} \\
\textbf{} & \textbf{} & \textbf{PS} & \textbf{-5.4\%} & \textbf{-7.3\%} & \textbf{-66.1\%} & \textbf{-10.9\%} & \textbf{-6.2\%} & \textbf{-7.4\%} & -91.4\% & \textbf{-11.0\%} & \textbf{-16.1\%} & \textbf{-76.0\%} & \textbf{-13.1\%} & \textbf{-15.4\%} & \underline{-19.6\%} & -81.9\% & \underline{-26.3\%} & \underline{-32.5\%} & \textbf{-81.2\%} & \underline{-32.8\%} & \underline{-29.3\%} & \underline{-31.9\%} & -88.7\% \\
\textbf{} & \textbf{} & \textbf{RJ} & \underline{-6.4\%} & \underline{-10.9\%} & -74.3\% & \underline{-15.7\%} & \underline{-9.1\%} & -12.5\% & -91.3\% & -22.2\% & -32.7\% & -83.9\% & -26.9\% & -26.8\% & -36.0\% & -86.2\% & -29.3\% & -35.5\% & -84.8\% & -36.7\% & -32.4\% & -34.7\% & \underline{-94.3\%} \\
\textbf{} & \textbf{} & \textbf{RW} & -7.4\% & -11.3\% & -74.9\% & -16.6\% & -9.8\% & -13.9\% & -90.2\% & -16.5\% & -27.6\% & -89.2\% & -17.6\% & -21.3\% & -39.0\% & \underline{-91.4\%} & -32.5\% & -40.2\% & \underline{-82.7\%} & -40.9\% & -36.2\% & -41.3\% & -91.4\% \\
\textbf{} & \textbf{} & \textbf{TS} & -9.8\% & -12.4\% & \underline{-69.9\%} & -17.7\% & -11.6\% & \underline{-11.5\%} & \underline{-91.8\%} & \underline{-12.3\%} & \underline{-16.6\%} & -87.3\% & \underline{-13.7\%} & \underline{-16.0\%} & \textbf{-18.9\%} & \textbf{-91.9\%} & \textbf{-22.7\%} & \textbf{-28.7\%} & -83.2\% & \textbf{-28.7\%} & \textbf{-25.3\%} & \textbf{-26.4\%} & -93.6\% \\
\cline{2-24}
\textbf{} & \multirow[c]{7}{*}{\textbf{0.20}} & \textbf{FF} & -5.8\% & -12.9\% & -71.5\% & -17.0\% & -11.7\% & -13.1\% & \textbf{-87.0\%} & -13.3\% & -19.3\% & -65.5\% & -17.8\% & -16.8\% & -20.6\% & -75.5\% & -20.5\% & -26.5\% & \underline{-66.7\%} & -26.8\% & -23.3\% & -27.3\% & -90.0\% \\
\textbf{} & \textbf{} & \textbf{FFB} & \underline{-3.6\%} & \underline{-6.6\%} & -54.2\% & \underline{-10.2\%} & \underline{-5.5\%} & -7.2\% & -85.2\% & -9.1\% & -11.3\% & -54.8\% & -9.9\% & -10.8\% & -11.9\% & -74.5\% & -18.8\% & -24.5\% & -74.7\% & -24.5\% & -21.7\% & -24.3\% & \textbf{-94.2\%} \\
\textbf{} & \textbf{} & \textbf{NS} & -3.8\% & -7.0\% & \textbf{18.4\%} & -10.9\% & -6.6\% & -10.5\% & -86.0\% & -39.2\% & -43.1\% & \textbf{54.1\%} & -40.8\% & -43.4\% & -45.2\% & \underline{-77.6\%} & -24.3\% & -31.5\% & -75.8\% & -31.6\% & -28.8\% & -31.9\% & \underline{-93.7\%} \\
\textbf{} & \textbf{} & \textbf{PS} & \textbf{-1.4\%} & \textbf{-1.1\%} & \underline{-36.1\%} & \textbf{-2.9\%} & \textbf{-0.7\%} & \textbf{-1.4\%} & -84.5\% & -6.9\% & -9.4\% & -56.6\% & -7.7\% & -9.7\% & -11.8\% & \textbf{-78.4\%} & \textbf{-15.5\%} & \underline{-20.2\%} & \textbf{-64.3\%} & \underline{-19.9\%} & \underline{-18.1\%} & \underline{-20.7\%} & -88.3\% \\
\textbf{} & \textbf{} & \textbf{RJ} & -3.7\% & -8.6\% & -60.7\% & -12.1\% & -7.5\% & -9.5\% & \underline{-86.8\%} & \underline{-6.2\%} & \underline{-6.3\%} & -55.1\% & \underline{-7.3\%} & \underline{-7.9\%} & \underline{-5.9\%} & -73.2\% & -22.2\% & -27.3\% & -71.8\% & -28.7\% & -25.2\% & -26.4\% & -92.6\% \\
\textbf{} & \textbf{} & \textbf{RW} & -3.6\% & -7.1\% & -57.7\% & -10.9\% & -6.0\% & -8.4\% & -84.7\% & \textbf{-3.8\%} & \textbf{-5.3\%} & -61.8\% & \textbf{-4.6\%} & \textbf{-6.2\%} & \textbf{-5.9\%} & -75.9\% & -18.3\% & -24.4\% & -67.8\% & -24.7\% & -21.0\% & -25.7\% & -90.7\% \\
\textbf{} & \textbf{} & \textbf{TS} & -7.0\% & -7.6\% & -47.7\% & -11.3\% & -6.9\% & \underline{-6.4\%} & -86.0\% & -14.2\% & -16.8\% & \underline{-53.3\%} & -15.7\% & -15.2\% & -16.4\% & -75.1\% & \underline{-15.5\%} & \textbf{-19.5\%} & -69.2\% & \textbf{-19.9\%} & \textbf{-17.7\%} & \textbf{-17.5\%} & -93.2\% \\
\cline{2-24}
\textbf{} & \multirow[c]{7}{*}{\textbf{0.50}} & \textbf{FF} & \underline{-0.4\%} & -3.0\% & -31.5\% & -4.0\% & -2.2\% & -2.2\% & -67.0\% & \underline{-1.8\%} & \underline{-0.6\%} & -24.9\% & -2.4\% & \underline{-1.4\%} & 0.4\% & -69.7\% & -6.6\% & -10.0\% & -41.6\% & -9.9\% & -7.7\% & -11.0\% & -89.5\% \\
\textbf{} & \textbf{} & \textbf{FFB} & \textbf{-0.4\%} & \textbf{-0.4\%} & -43.8\% & \textbf{-1.0\%} & \textbf{0.0\%} & \underline{-0.2\%} & \underline{-70.9\%} & -2.1\% & -1.0\% & \underline{-24.0\%} & \underline{-1.5\%} & -2.0\% & 1.0\% & -73.3\% & \textbf{-4.4\%} & \underline{-6.0\%} & \textbf{-19.3\%} & \textbf{-6.0\%} & \underline{-5.2\%} & -6.2\% & -86.2\% \\
\textbf{} & \textbf{} & \textbf{NS} & -1.8\% & -4.0\% & \textbf{37.2\%} & -6.2\% & -4.3\% & -7.1\% & -69.1\% & -19.1\% & -20.1\% & \textbf{40.1\%} & -20.6\% & -22.4\% & -19.7\% & \textbf{-85.1\%} & -8.5\% & -11.9\% & -40.6\% & -11.2\% & -10.5\% & -11.0\% & \textbf{-92.4\%} \\
\textbf{} & \textbf{} & \textbf{PS} & -0.8\% & \underline{-0.4\%} & -30.8\% & \underline{-1.7\%} & \underline{-0.0\%} & \textbf{0.4\%} & -70.8\% & -5.6\% & -6.5\% & -29.1\% & -6.4\% & -6.2\% & -5.6\% & -75.1\% & -4.6\% & -6.3\% & -28.3\% & \underline{-6.4\%} & -5.2\% & \textbf{-5.4\%} & -87.5\% \\
\textbf{} & \textbf{} & \textbf{RJ} & -0.6\% & -2.0\% & \underline{-27.0\%} & -3.1\% & -1.5\% & -2.2\% & -65.8\% & \textbf{-0.4\%} & \textbf{1.2\%} & -34.7\% & \textbf{-0.3\%} & \textbf{-0.8\%} & \underline{2.3\%} & \underline{-80.1\%} & -5.2\% & -7.1\% & \underline{-24.0\%} & -7.4\% & -5.9\% & -7.5\% & -85.4\% \\
\textbf{} & \textbf{} & \textbf{RW} & -1.5\% & -3.2\% & -27.9\% & -3.8\% & -3.1\% & -2.8\% & -63.6\% & -3.7\% & -0.8\% & -26.1\% & -4.3\% & -3.5\% & \textbf{2.4\%} & -64.7\% & \underline{-4.6\%} & \textbf{-5.8\%} & -27.6\% & -7.2\% & \textbf{-4.8\%} & \underline{-5.5\%} & -85.1\% \\
\textbf{} & \textbf{} & \textbf{TS} & -5.1\% & -5.6\% & -39.5\% & -8.1\% & -4.6\% & -4.6\% & \textbf{-73.7\%} & -10.4\% & -10.6\% & -25.6\% & -12.3\% & -10.6\% & -8.3\% & -78.7\% & -5.8\% & -7.2\% & -31.8\% & -7.3\% & -6.9\% & -6.8\% & \underline{-89.8\%} \\
\cline{1-24} \cline{2-24}
\multirow[c]{27}{*}{\textbf{PinSAGE}} & \textbf{-} & \textbf{-} & \textbf{0.583} & \textbf{0.152} & \textbf{0.039} & \textbf{0.065} & \textbf{0.131} & \textbf{0.086} & \textbf{1.071} & \textbf{0.122} & \textbf{0.027} & \textbf{0.159} & \textbf{0.054} & \textbf{0.008} & \textbf{0.012} & \textbf{3.036} & \textbf{0.136} & \textbf{0.024} & \textbf{0.060} & \textbf{0.041} & \textbf{0.009} & \textbf{0.011} & \textbf{11.337} \\
\cline{2-24}
\textbf{} & \multirow[c]{6}{*}{\textbf{0.05}} & \textbf{FF} & \underline{10.7\%} & -17.8\% & -51.4\% & 4.4\% & -17.0\% & -31.7\% & \textbf{-94.8\%} & -73.0\% & -75.6\% & -69.7\% & -74.1\% & -74.6\% & -70.7\% & -94.9\% & -73.4\% & -78.0\% & -82.3\% & -77.7\% & -75.8\% & -79.5\% & -93.1\% \\
\textbf{} & \textbf{} & \textbf{FFB} & -4.2\% & -21.2\% & -35.4\% & -18.0\% & -14.6\% & -32.2\% & -88.7\% & -48.9\% & -59.9\% & \underline{15.8\%} & -56.5\% & -53.7\% & -57.5\% & -88.1\% & \underline{-53.4\%} & \underline{-59.8\%} & 2.9\% & \underline{-62.0\%} & \underline{-57.4\%} & \underline{-62.4\%} & \underline{-93.1\%} \\
\textbf{} & \textbf{} & \textbf{PS} & 10.5\% & \textbf{4.9\%} & \textbf{-6.1\%} & \textbf{13.0\%} & \textbf{0.9\%} & \textbf{-0.0\%} & -84.5\% & -64.4\% & -72.6\% & \textbf{87.0\%} & -68.5\% & -67.9\% & -66.6\% & -94.8\% & -58.9\% & -66.0\% & \textbf{76.2\%} & -66.2\% & -64.0\% & -65.2\% & -90.1\% \\
\textbf{} & \textbf{} & \textbf{RJ} & -11.3\% & -15.2\% & -30.4\% & -19.1\% & -14.2\% & -25.1\% & -88.1\% & \underline{-45.0\%} & -49.9\% & -25.2\% & -52.4\% & \underline{-47.6\%} & \underline{-41.9\%} & \underline{-95.1\%} & -58.4\% & -64.3\% & 19.7\% & -65.4\% & -62.4\% & -66.5\% & \textbf{-94.1\%} \\
\textbf{} & \textbf{} & \textbf{RW} & 3.5\% & \underline{-2.1\%} & -35.4\% & 5.2\% & -2.8\% & -20.3\% & \underline{-91.1\%} & -48.4\% & \underline{-48.5\%} & -33.3\% & \underline{-52.3\%} & -53.9\% & -42.4\% & -92.7\% & -70.7\% & -76.2\% & -30.8\% & -76.6\% & -73.4\% & -78.9\% & -92.8\% \\
\textbf{} & \textbf{} & \textbf{TS} & \textbf{13.2\%} & -2.6\% & \underline{-21.5\%} & \underline{8.3\%} & \underline{-2.4\%} & \underline{-10.5\%} & -88.0\% & \textbf{-20.6\%} & \textbf{-28.8\%} & -13.1\% & \textbf{-25.6\%} & \textbf{-26.3\%} & \textbf{-27.0\%} & \textbf{-96.1\%} & \textbf{-41.2\%} & \textbf{-47.7\%} & \underline{61.0\%} & \textbf{-48.7\%} & \textbf{-44.6\%} & \textbf{-53.5\%} & -85.5\% \\
\cline{2-24}
\textbf{} & \multirow[c]{6}{*}{\textbf{0.10}} & \textbf{FF} & \textbf{14.6\%} & \textbf{4.3\%} & -29.3\% & \textbf{24.3\%} & \textbf{3.4\%} & -15.8\% & -88.3\% & -51.5\% & -63.7\% & -54.4\% & -57.1\% & -57.3\% & -63.5\% & -94.4\% & -66.3\% & -72.3\% & -87.2\% & -72.8\% & -69.3\% & -75.2\% & \textbf{-93.1\%} \\
\textbf{} & \textbf{} & \textbf{FFB} & -5.5\% & -8.5\% & -26.0\% & -10.3\% & -7.2\% & -19.6\% & -76.2\% & -28.8\% & \textbf{-24.4\%} & \underline{32.4\%} & -32.9\% & -29.7\% & \textbf{-8.9\%} & -81.5\% & \underline{-37.8\%} & \underline{-44.2\%} & \textbf{54.8\%} & \underline{-45.8\%} & \underline{-42.2\%} & -49.3\% & -83.9\% \\
\textbf{} & \textbf{} & \textbf{PS} & 4.4\% & \underline{-1.4\%} & \textbf{-9.9\%} & 4.2\% & \underline{0.7\%} & \textbf{-3.2\%} & -76.2\% & \underline{-25.5\%} & -31.8\% & \textbf{64.3\%} & \textbf{-27.6\%} & \underline{-28.9\%} & -34.7\% & \underline{-95.5\%} & -42.6\% & -48.7\% & 25.3\% & -50.7\% & -48.4\% & \underline{-47.7\%} & -85.2\% \\
\textbf{} & \textbf{} & \textbf{RJ} & -5.6\% & -8.8\% & -27.6\% & -9.9\% & -6.8\% & -20.0\% & \underline{-88.5\%} & -39.4\% & -47.9\% & -25.5\% & -44.5\% & -43.0\% & -48.1\% & -95.3\% & -47.5\% & -52.4\% & \underline{47.6\%} & -55.4\% & -51.5\% & -53.7\% & -87.9\% \\
\textbf{} & \textbf{} & \textbf{RW} & \underline{9.3\%} & -4.9\% & -35.9\% & \underline{7.9\%} & -6.6\% & -21.4\% & \textbf{-90.6\%} & -48.6\% & -56.7\% & -17.0\% & -53.5\% & -50.2\% & -54.3\% & -94.3\% & -52.0\% & -57.5\% & 21.5\% & -59.5\% & -55.1\% & -60.6\% & \underline{-89.1\%} \\
\textbf{} & \textbf{} & \textbf{TS} & -1.3\% & -9.3\% & \underline{-15.5\%} & -12.0\% & -8.3\% & \underline{-10.1\%} & -88.2\% & \textbf{-24.8\%} & \underline{-29.0\%} & -15.0\% & \underline{-28.1\%} & \textbf{-28.1\%} & \underline{-27.6\%} & \textbf{-95.5\%} & \textbf{-31.4\%} & \textbf{-36.8\%} & 29.5\% & \textbf{-39.8\%} & \textbf{-35.6\%} & \textbf{-38.8\%} & -84.3\% \\
\cline{2-24}
\textbf{} & \multirow[c]{7}{*}{\textbf{0.20}} & \textbf{FF} & -1.5\% & -15.3\% & -21.0\% & -7.6\% & -13.5\% & -20.5\% & \textbf{-87.7\%} & -26.6\% & -33.7\% & -36.5\% & -31.4\% & -28.6\% & -32.4\% & -85.3\% & -41.7\% & -50.7\% & -58.6\% & -50.7\% & -45.7\% & -53.6\% & \underline{-89.4\%} \\
\textbf{} & \textbf{} & \textbf{FFB} & \textbf{11.3\%} & \textbf{-0.0\%} & -9.9\% & \textbf{15.0\%} & 0.5\% & -10.7\% & -59.4\% & -27.3\% & -29.4\% & 6.0\% & -29.1\% & -28.8\% & \underline{-24.5\%} & -83.6\% & \underline{-30.2\%} & \underline{-34.5\%} & 59.0\% & \underline{-36.0\%} & \underline{-34.2\%} & \underline{-36.4\%} & -64.7\% \\
\textbf{} & \textbf{} & \textbf{NS} & 1.8\% & -32.7\% & -52.5\% & -20.7\% & -36.3\% & -48.8\% & \underline{-87.7\%} & -46.0\% & -44.6\% & \textbf{199.6\%} & -47.8\% & -46.2\% & -38.5\% & -86.0\% & -82.1\% & -86.5\% & \underline{78.5\%} & -85.5\% & -85.4\% & -86.8\% & -70.1\% \\
\textbf{} & \textbf{} & \textbf{PS} & -3.9\% & -3.4\% & \textbf{3.9\%} & -5.6\% & -2.9\% & \underline{-4.0\%} & -51.7\% & -25.5\% & -31.6\% & \underline{29.7\%} & -32.0\% & -29.1\% & -26.9\% & \underline{-87.3\%} & -35.5\% & -42.3\% & 16.9\% & -41.3\% & -39.3\% & -42.9\% & -81.6\% \\
\textbf{} & \textbf{} & \textbf{RJ} & -4.9\% & -4.0\% & -16.0\% & -5.3\% & -1.6\% & -13.3\% & -74.0\% & \underline{-17.8\%} & \underline{-26.5\%} & 14.8\% & \underline{-22.9\%} & \underline{-23.5\%} & -24.9\% & -82.0\% & -41.1\% & -47.2\% & 19.6\% & -49.6\% & -45.3\% & -49.1\% & \textbf{-89.6\%} \\
\textbf{} & \textbf{} & \textbf{RW} & \underline{8.2\%} & \underline{-0.6\%} & -17.7\% & \underline{14.8\%} & \textbf{0.5\%} & -12.0\% & -79.1\% & -45.5\% & -56.7\% & -40.9\% & -52.0\% & -50.2\% & -50.0\% & \textbf{-92.5\%} & -34.6\% & -41.9\% & \textbf{112.0\%} & -42.4\% & -37.3\% & -46.8\% & -69.8\% \\
\textbf{} & \textbf{} & \textbf{TS} & 6.7\% & -0.9\% & \underline{-8.3\%} & 3.9\% & \underline{0.5\%} & \textbf{-1.6\%} & -59.3\% & \textbf{-15.5\%} & \textbf{-21.0\%} & 9.5\% & \textbf{-20.3\%} & \textbf{-18.1\%} & \textbf{-17.9\%} & -86.1\% & \textbf{-20.5\%} & \textbf{-25.3\%} & 35.5\% & \textbf{-25.9\%} & \textbf{-23.7\%} & \textbf{-29.1\%} & -73.0\% \\
\cline{2-24}
\textbf{} & \multirow[c]{7}{*}{\textbf{0.50}} & \textbf{FF} & -3.3\% & -3.2\% & -6.6\% & -3.1\% & 0.3\% & -5.7\% & \textbf{-55.0\%} & -20.7\% & -30.7\% & -22.6\% & -27.1\% & -23.0\% & -27.5\% & \textbf{-74.0\%} & -18.3\% & -21.8\% & 56.5\% & -22.7\% & -21.0\% & -22.6\% & \underline{-41.3\%} \\
\textbf{} & \textbf{} & \textbf{FFB} & \underline{9.2\%} & \underline{6.1\%} & -5.0\% & \underline{15.3\%} & \underline{5.9\%} & -0.8\% & 2.9\% & -18.4\% & -19.7\% & 15.0\% & -20.6\% & -18.3\% & -13.4\% & -55.6\% & \textbf{-14.8\%} & \textbf{-17.3\%} & 42.9\% & \underline{-19.4\%} & \textbf{-17.6\%} & \textbf{-19.0\%} & -6.1\% \\
\textbf{} & \textbf{} & \textbf{NS} & -21.0\% & -50.5\% & \textbf{2.2\%} & -57.2\% & -48.3\% & -52.7\% & \underline{-35.8\%} & -41.6\% & -46.0\% & \textbf{62.7\%} & -46.8\% & -44.1\% & -39.6\% & -58.5\% & -70.1\% & -75.1\% & \underline{77.2\%} & -74.1\% & -75.4\% & -74.4\% & -6.7\% \\
\textbf{} & \textbf{} & \textbf{PS} & \textbf{13.2\%} & \textbf{6.4\%} & \textbf{2.2\%} & \textbf{21.2\%} & \textbf{6.1\%} & \textbf{2.0\%} & -14.3\% & -10.2\% & \underline{-13.8\%} & \underline{25.1\%} & -14.0\% & -11.6\% & \underline{-10.3\%} & -4.0\% & -16.6\% & -19.3\% & 52.7\% & \textbf{-19.1\%} & -19.7\% & \underline{-20.2\%} & -16.3\% \\
\textbf{} & \textbf{} & \textbf{RJ} & 4.8\% & 2.0\% & -3.3\% & 7.3\% & 3.5\% & -2.8\% & -26.1\% & \underline{-9.1\%} & -14.0\% & 12.6\% & \underline{-12.1\%} & \underline{-9.2\%} & -16.1\% & -49.4\% & \underline{-15.5\%} & \underline{-18.6\%} & 71.4\% & -19.9\% & \underline{-17.8\%} & -20.4\% & -18.5\% \\
\textbf{} & \textbf{} & \textbf{RW} & -0.1\% & 0.4\% & -3.9\% & 0.5\% & 0.7\% & -3.8\% & -6.4\% & -12.5\% & -15.9\% & 8.9\% & -15.1\% & -11.0\% & -10.4\% & -50.9\% & -16.7\% & -20.4\% & \textbf{96.5\%} & -20.7\% & -18.7\% & -24.1\% & -21.2\% \\
\textbf{} & \textbf{} & \textbf{TS} & 7.5\% & 2.8\% & -3.9\% & 5.9\% & 2.9\% & \underline{1.4\%} & -34.7\% & \textbf{-8.3\%} & \textbf{-11.2\%} & 9.5\% & \textbf{-10.7\%} & \textbf{-8.0\%} & \textbf{-9.8\%} & \underline{-65.8\%} & -19.1\% & -23.8\% & -33.3\% & -24.1\% & -22.5\% & -25.8\% & \textbf{-70.1\%} \\
\cline{1-24} \cline{2-24}
\end{tabular}